\documentclass[conference]{IEEEtran}
\usepackage{amsmath,amssymb,amsfonts}
\usepackage{pifont}
\usepackage{algorithmic}
\usepackage{subfig}
\usepackage{graphicx}
\usepackage{textcomp}
\usepackage{xcolor}
\usepackage{fancyhdr}
\usepackage[hyphens]{url}
\usepackage[sort,nocompress]{cite}
\usepackage[final]{microtype}
\usepackage[keeplastbox]{flushend}
\usepackage{multirow}
\usepackage{latexsym}
\usepackage{scalefnt}
\usepackage{comment}
\usepackage[bookmarks=true,breaklinks=true,letterpaper=true,colorlinks,linkcolor=black,citecolor=blue,urlcolor=black]{hyperref}
\def\BibTeX{{\rm B\kern-.05em{\sc i\kern-.025em b}\kern-.08em
    T\kern-.1667em\lower.7ex\hbox{E}\kern-.125emX}}
\def \debug{}

\ifx \debug \undefined
\newcommand{\am}[1]{{}} 
\newcommand{\pn}[1]{{}} 
\newcommand{\mz}[1]{{}} 
\else
\newcommand{\am}[1]{{\textcolor{blue}{AM: #1}}} 
\newcommand{\pn}[1]{{\textcolor{magenta}{PN: #1}}} 
\newcommand{\mz}[1]{{\textcolor{red}{MZ: #1}}} 
\fi

\newcommand{\scheme}{\textsc{Forecaster}}
\title{The Case for Learning Application Behavior to Improve Hardware Energy Efficiency}
\author{
\IEEEauthorblockN{Kevin Weston, Vahid Janfaza, Abhishek Taur, Abdullah Muzahid}
\IEEEauthorblockA{\textit{Texas A\&M University}}
\and
\IEEEauthorblockN{Arnav Kansal, Mohamed Zahran}
\IEEEauthorblockA{\textit{New York University}}
}

\begin{document}
\maketitle
\pagestyle{plain}









\begin{abstract}
Computer applications are continuously evolving.
However, significant knowledge can be harvested from a set of applications and applied in the context of unknown applications. 
In this paper, we propose to use the harvested knowledge to tune hardware configurations.
The goal of such tuning is to maximize hardware efficiency (i.e., maximize an application's performance while minimizing the energy consumption). Our proposed approach, called \scheme, uses a deep learning model to learn what configuration of hardware resources provides the optimal energy efficiency for a certain behavior of an application. During the execution of an unseen application, the model uses the learned knowledge to reconfigure hardware resources in order to maximize energy efficiency.
We have provided a detailed design and implementation of \scheme\ and compared its performance against a prior state-of-the-art hardware reconfiguration approach. Our results show that \scheme\ can save as much as 18.4\% system power over the baseline set up with all resources. On average, \scheme\ saves 16\% system power over the baseline setup while sacrificing less than 0.01\% of overall performance.
Compared to the prior scheme, \scheme\ increases power savings by 7\%.




\end{abstract}

\section{Introduction}
\label{sec:intro}
The landscape of computing applications is continuously evolving. Even within the last 20 years, we have seen a broad spectrum of application types such as single-threaded, multi-threaded, desktop-based, cloud-based, portable device-based, machine learning-oriented, etc~\cite{swhistory}. The lesson to take from such an evolving domain is that computing systems should not be designed or built with a predetermined set of applications in mind. Rather a computing system should be nimble enough to adapt itself with any unknown application that may come in its way.

We propose to use {\em application behavior} learning as a general principle to cope up with ever-changing
computing applications. We argue that a computing system should
learn the behavior of a diverse set of applications. 
The intuition is that although different applications have different functionalities, they have a commonality in behavior to a certain extent. Therefore, we can learn what works (or does not work) in achieving a certain goal from a set of applications and apply the accumulated {\em knowledge/experience} to different or unknown applications to achieve similar goals. This knowledge sharing is the core idea behind application behavior learning.
As a case study, we investigate in this paper whether application behavior learning can be applied to improve the energy efficiency of computer hardware.
Computer architects are in a continuous quest to find the best hardware design to improve energy efficiency without sacrificing performance.
It is not feasible to have program specific hardware designs for different program types.
To make things worse, a single program is found to go through different distinct phases during its execution~\cite{sherwood02, sherwood03} 
where each phase requires a different hardware configuration to achieve an optimal energy-performance trade-off~\cite{dubach10}.
This paper outlines a promising solution for this problem using application behavior learning as the guiding principle. 

The usual practice is to
gather profiling information about program executions on a hardware, and then make use of the profiling information to either enhance the hardware or the program~\cite{dubach10, ipek08}.
However, this implies that each program must be instrumented and profiled first, and the profiled information is used for that application only.
We hypothesize that {\em different phases of an application can have similarities with those of another application} (Section~\ref{sec:motiv}). 
Since a program phase requires a specific configuration for the optimal energy efficiency, 
if there is a different program with a phase that has similar behavior, then we can use the same configuration to get the optimal energy efficiency. 
Thus, the task would be then to {\em learn} the best configuration for different program phases from a set of diverse applications 
and use the accumulated knowledge to determine the best configuration for new, unknown programs. 
{\em The goal of this paper is to develop and implement such a technique inside the hardware.} 

There are two major challenges towards this goal.
{\em First,} which hardware structures should be reconfigured?
There are many structures that can be designed to be reconfigurable. 
We choose the structures that have the biggest impact on performance and power, and at the same time 
can be reconfigured with the least hardware cost and modification.
{\em Second,} how to learn the patterns of the hardware/software interaction to choose the best configuration?
We choose to periodically collect hardware telemetry (collection of data points) from various hardware performance counters.
However, the counters along with different phase behavior can lead to an intractable number of possibilities to
explore before we can determine the best hardware configuration. That is why, a naive classifier using bloom filter, or regression analysis
may not be effective~\cite{beyondprofiling}. Instead, we choose to use a deep neural network which has been improving steadily over the last few years~\cite{imagenet, vgg16, lstm} and shown to provide incredible accuracy for a number of challenging tasks such as image recognition, language translation, speech recognition, or autonomous control of vehicles.

We propose \scheme\ to address the challenges.
\scheme\ periodically collects hardware telemetry during the execution of a program. \scheme\ uses the hardware telemetry in a deep learning model to predict the configuration of tunable hardware resources
to maximize the energy efficiency without degrading the application's performance significantly. 
We provide a detailed design and implementation of \scheme\ using Multi2Sim~\cite{multi2sim} simulator.
Our experimental results using PARSEC benchmarks show that the proposed technique can save 
as much as 18.4\% power over the baseline with all resources. On average, our scheme saves
16\% system power over the baseline setup 
while sacrificing 0.007\% of overall performance.
Compared to a prior state-of-the-art scheme~\cite{dubach10}, \scheme\ increases power savings by 7\%.


\section{Motivation}
\label{sec:motiv}
\scheme\ is grounded on two simple observations - {\em (i) there are significant similarities in execution phases across 
applications, and (ii) each execution phase requires a specific hardware configuration to maximize energy efficiency 
without hurting performance}. 

\begin{figure}[htpb]
    \centering
    
    {\includegraphics[width=\columnwidth]{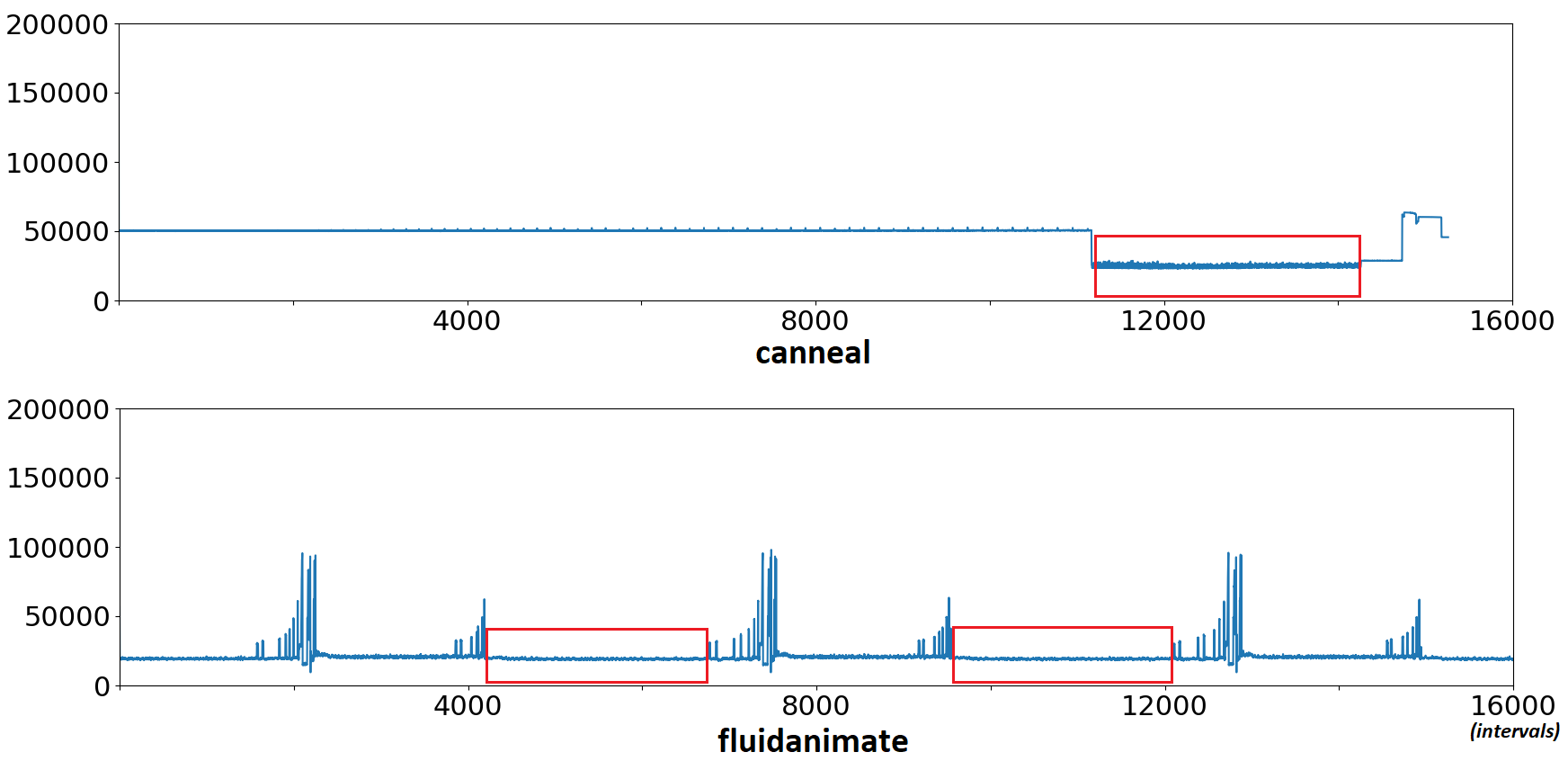}}
    \caption{Number of branch predictions per interval of 0.5 million instructions. Similarities are highlighted in colored boxes. \label{fig-observe1-branch}}

    {\includegraphics[width=\columnwidth]{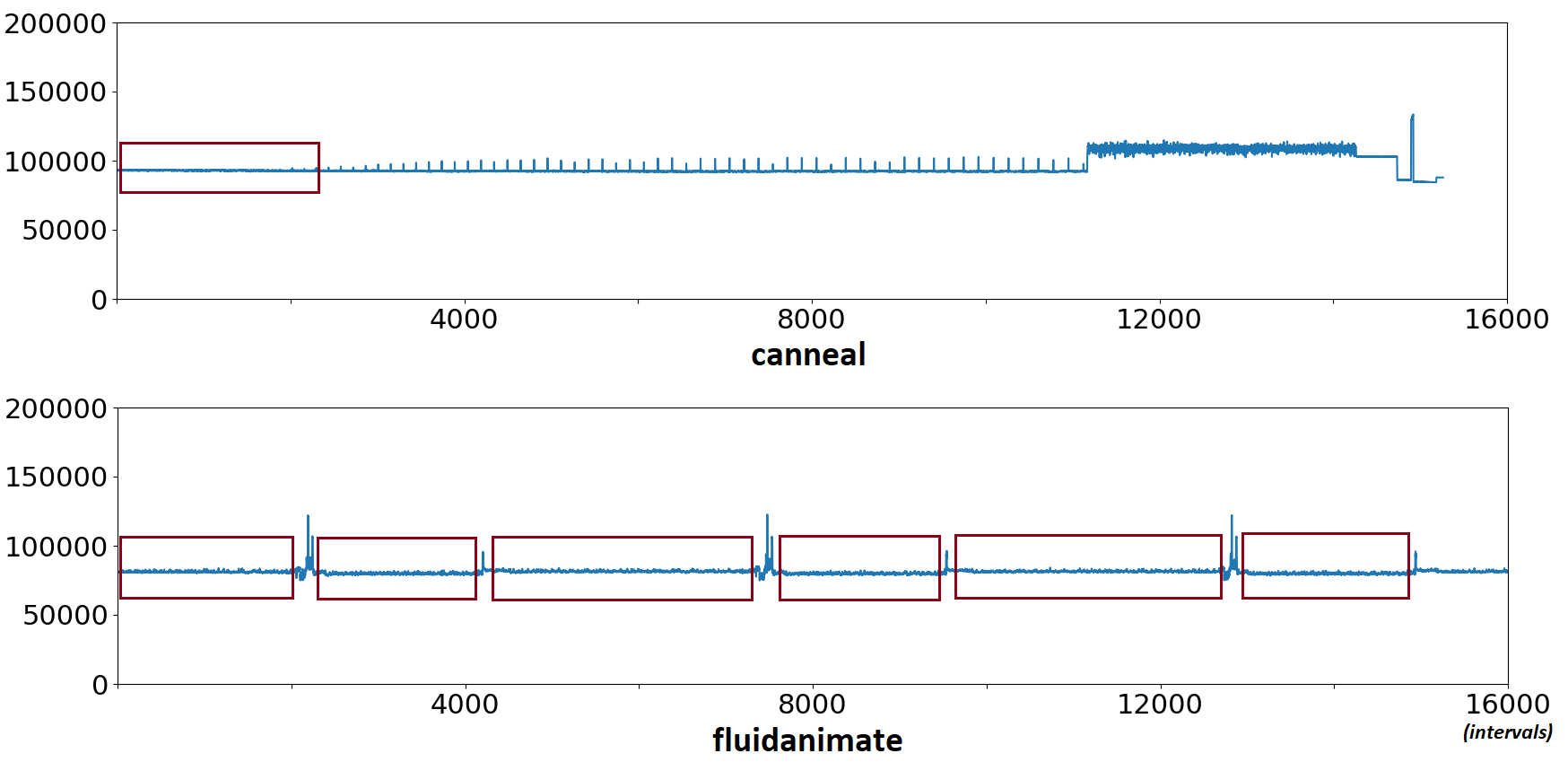}}  
    \caption{Number of L1-data accesses per interval of 0.5 million instructions. Similarities are highlighted in colored boxes. \label{fig-observe1-dl1}}
    \vspace{-0.2cm}
    
\end{figure}

In order to support the first observation, we analyze two applications 
- \textit{canneal} and \textit{fluidanimate} from Parsec. Figure~\ref{fig-observe1-branch} \& \ref{fig-observe1-dl1} show the the number of branch predictions and L1-data accesses in every interval of 0.5 million instructions. The red colored boxes in Figure~\ref{fig-observe1-branch} shows that one execution phase of \textit{canneal} is similar to two execution phases of \textit{fluidanimate} where the number of branch predictions is steady at around 20,000 per interval. Thus, the control flow structure of these execution phases between two different applications are similar. If we consider L1-data accesses, Figure~\ref{fig-observe1-dl1} shows that one execution phase of \textit{canneal} is similar to 6 other execution phases of \textit{fluidanimate}. Therefore, the data access patterns of these phases of the applications should be similar too. In other words, despite being two completely different applications with different functionalities, \textit{canneal} and \textit{fluidanimate} share a lot of similarities among their execution phases.


\begin{figure}[htpb]
\centering
\includegraphics[width=\columnwidth]{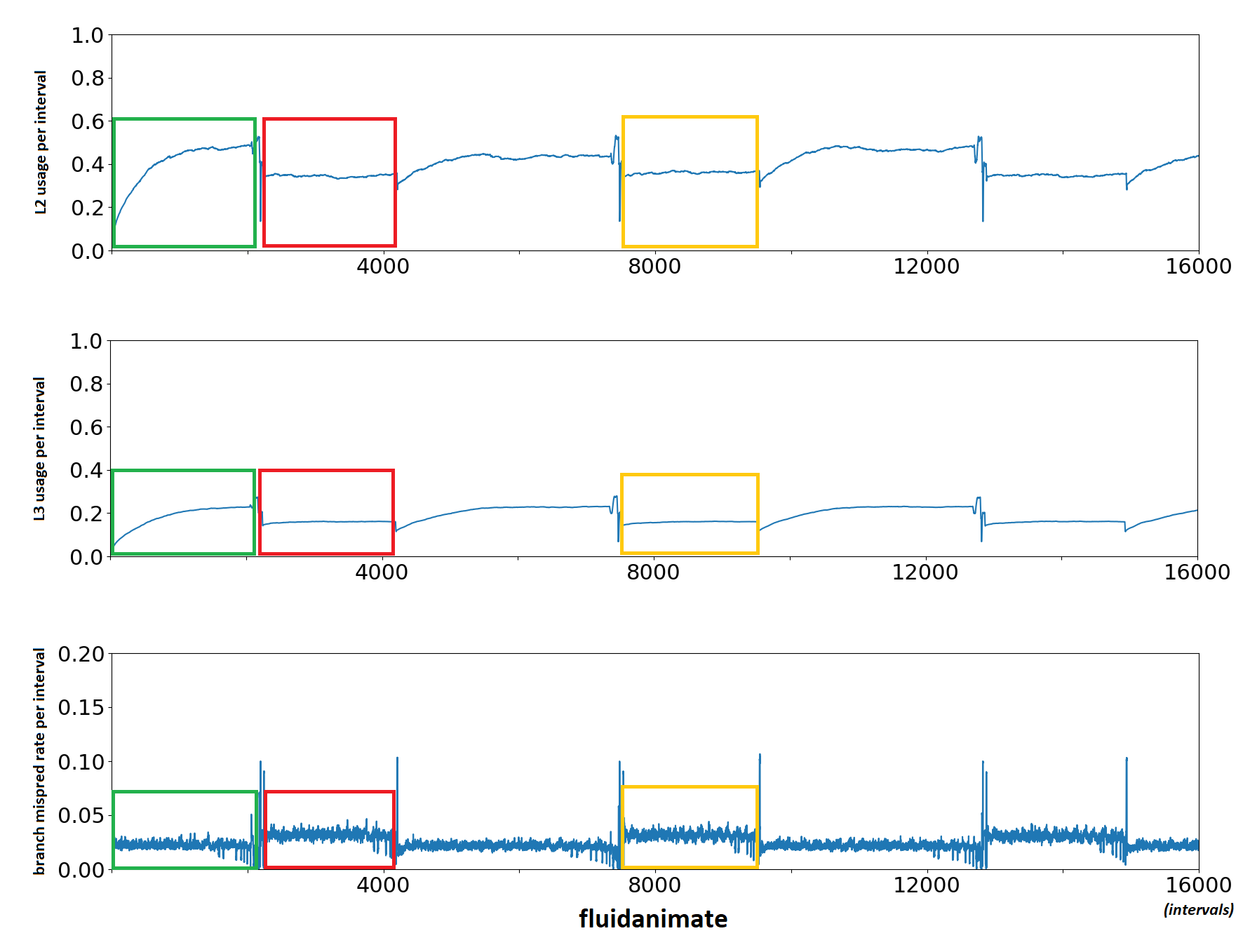}
\caption{Time-series data of branch misprediction rate, L2 and L3 usage of \textit{fluidanimate} during execution.}
\label{fig-observe2} 
 \vspace{-0.2cm}
\end{figure}

Figure~\ref{fig-observe2} shows the detailed time-series characteristics of L2 and L3 usage as well as branch mispredictions of \textit{fluidanimate}. It shows that the first execution phase (shown in green boxes) has different L2, L3 access and branch characteristics than the second phase (shown in red boxes). Therefore, the hardware configuration that provides the optimal trade-off between energy efficiency and performance for the first phase is different than that of the second phase. For example, since the first phase uses about 45\% of L2 and 25\% of L3, an optimal cache configuration of the first phase consists of 60\% of L2 and 40\% of L3. Similarly, the optimal configuration for the second phase is a combination of 40\% of L2 and 20\% of L3. This clearly demonstrates that every distinct phase requires a different hardware configuration to maximize energy efficiency while maintaining performance. Note that the forth phase (in yellow box) is quite similar to the second phase and therefore, require the same hardware configuration as the second phase.

\section{Background \& Related Work}
\label{sec:back}

There is a considerable amount of prior work on reconfigurable architecture~\cite{dubach10, bala00, flicker, wildstrom07, bitirgen08, samplepgo, tarsa19, ipek08, charstar}, which can be grouped into two categories depending on their implementation: hardware-based and software-based techniques.

\subsection{Hardware-based Adaptive Architecture}
Work in this category propose the use of dedicated hardware chip to host the optimization module. 
Dubach et al.~\cite{dubach10} propose the use of machine learning to dynamically optimize the efficiency of some processor's components such as the Arithmetic Logic Unit, instruction queues, register file, caches, branch predictor, and the pipeline depth. During program execution, as soon as a phase change is detected, the hardware starts to collect counters on a predefined profiling configuration. These counters represent the usage of the hardware resources in that interval. The model then predicts the optimal configuration and the system is reconfigured accordingly for the rest of the phase. They propose using simple bitline segmentation technique to make processor structure adaptive. Their predictive model is stored on a separate chip and requires about 2KB of storage.

Choi and Yeung~\cite{choi06} perform microarchitectural resources distribution in an SMT processor using hill-climbing algorithm. Bitirgen et al.~\cite{bitirgen08} propose a scheme to combine performance prediction model of multiple applications to get an aggregate performance prediction of the overall resource distribution. The scheme is coupled with some limited probabilistic search technique to find the optimal resource distribution to improve performance. Petrica et al.~\cite{flicker} present Flicker, a general-purpose multicore architecture that dynamically adapts to varying limits on allocated power. A Flicker core has reconfigurable lanes through the pipeline that allows tailoring an individual core to the running application with lower overhead. Ravi et al.~\cite{charstar} propose CHARSTAR, a clock tree aware resource optimizing mechanism. CHARSTAR incorporates a multi-layer perceptron with one hidden layer to predict the optimal configuration in each execution phase. The neural network takes into account the clock hierarchy and the topology overhead in order to improve the power savings. Secondly, CHARSTAR only works for single-threaded programs, and a multi-threaded version may cause a super-linearly increase in the size of the neural network model.

\subsection{Software-based Adaptive Architecture}
Optimization techniques in this category are implemented in software~\cite{samplepgo, tarsa19}. They are usually a part of the operating system, compiler, or microcontroller firmware. Tarsa et al.~\cite{tarsa19} propose a lightweight ML framework that can be distributed through firmware updates to the microcontroller for post-silicon CPUs. The ML model is first trained offline with a diverse collection of applications to avoid statistical blind spots. During execution, the CPU dynamically sets the issue width of a clustered hardware component while clock-gating unused resources based on the prediction of the ML model.

There is also a well-established line of work that tries to achieve energy-performance trade-off without any hardware structural adaptation. Prominent works that fall in this category use dynamic voltage-frequency scaling (DVFS)~\cite{dvfs,sysscale, coscale}. DVFS is an approach that dynamically adjust the frequency and voltage of CPU components to increase system power efficiency. Under ideal circumstance, DVFS can save a quadratic factor of dynamic power while sacrificing only a linear factor of overall performance. However, applying this technique in real world systems can be tricky because reduced frequency means longer execution time. Thus, an inefficient DVFS algorithm may unintentionally increase the overall energy consumption. Haj-Yahya et al. propose SysScale~\cite{sysscale}, a power management scheme that control the power budget across IO, computing, and memory domains. SysScale uses a DVFS mechanism to allocate power budget to each domain based on the predicted demands, results in an improvement of 16\% in CPU workload and 8.9\% in graphical-related workload. The implementation of SysScale requires a modification of the firmware to support the transition of different frequency/voltage operating points.

\begin{figure*}[hptb]
\centering
\includegraphics[width=0.7\linewidth]{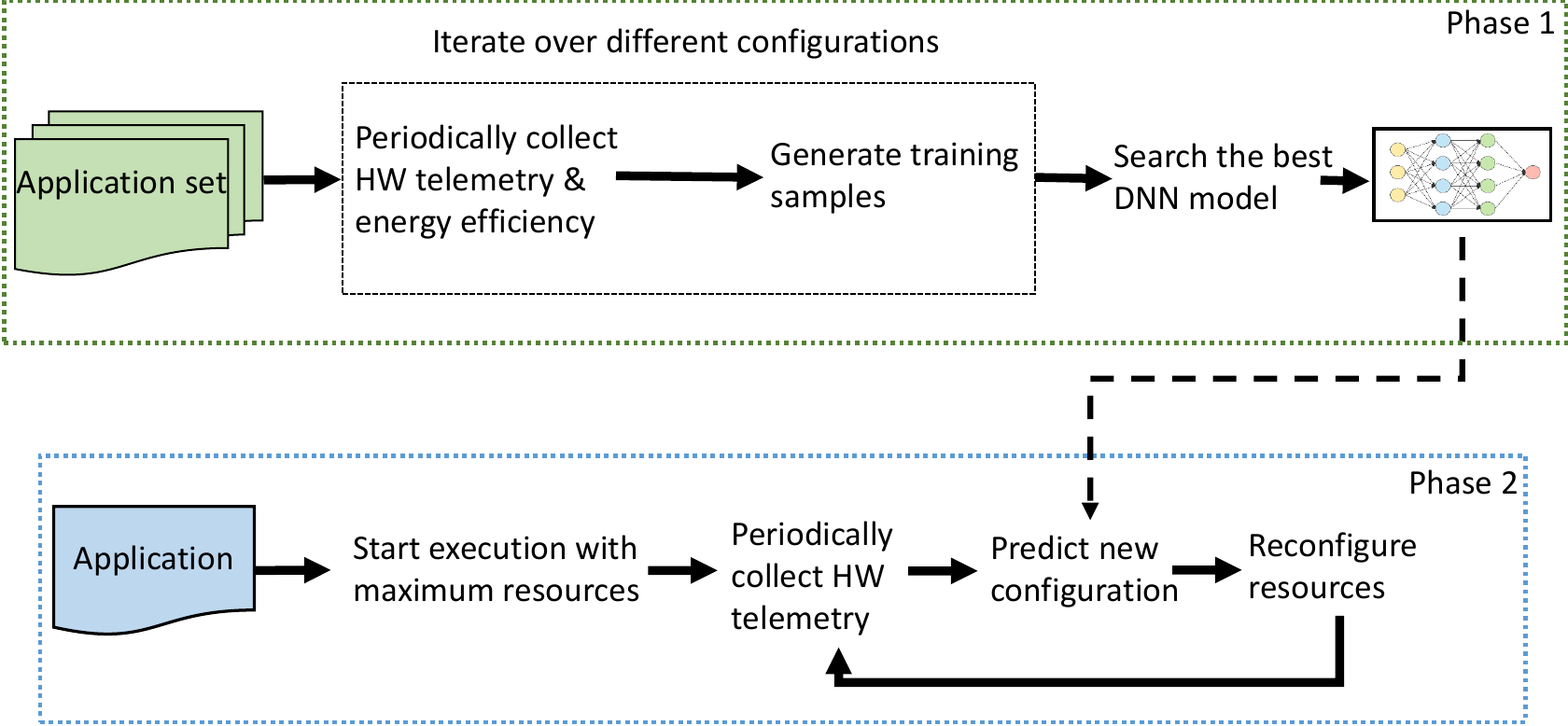}
\caption{Overall workflow of \scheme.}
\vspace{-0.5cm}
\label{fig-workflow} 
\end{figure*}

\section{Main Idea: \scheme}
\label{sec:desc}

\scheme\ works in two phases - (i) building a model that is able to predict the best configuration of hardware resources for maximizing
the energy efficiency (i.e., $Instruction~Per~Second~(IPS)^3/Power$) and (ii) changing the hardware resources according to the predicted configuration. 
Figure~\ref{fig-workflow} shows the overall workflow. \scheme\ works in the first phase 
only once using a set of applications whereas the second phase happens at runtime repeatedly during the execution of any application.
Both phases use hardware telemetry collected during the execution of an application.
The telemetry consists of various hardware event counters
that implicitly capture the behavior of the application. The first phase uses the telemetry to build a dataset which is used to 
train a deep neural network (DNN) model. The second phase uses the trained DNN model to predict and reconfigure hardware
resources. The following sections elaborate on both phases.

\subsection{Phase 1: Building a Predictive Model}
\label{sec-model}

\scheme\ builds a predictive model by first collecting hardware telemetry on a set of benchmarks for different configurations of hardware resources and then, training a DNN model on the dataset. 

\subsubsection{Selecting Hardware Resources}
\label{sec-res-sel}
As reconfigurable hardware resources, we choose L2 and L3 caches as well as the Branch Target Buffer (BTB) and Prefetcher. We choose caches because they are the most energy hungry resources in a modern chip~\cite{cache-energy}. We choose the other resources because they can be easily clock-gated without intrusive changes to the pipeline circuitry (Section~\ref{sec-reconfig}). Although we demonstrate the effectiveness of \scheme\ with these 4 resources in this paper, we argue that \scheme\ is general enough to accommodate any number of resources. Table~\ref{table-knob} shows the reconfigurable resources and possible configurations. We used $IPS^3/Power$ as the metric to calculate energy efficiency. The same metric has been used in prior work too~\cite{dubach10}.

\begin{table}[h!]
\centering
\scalebox{0.8}{
\begin{tabular}{||l| l ||} 
 \hline\hline
 \multicolumn{1}{||c|}{Tunable Resource} & \multicolumn{1}{c||}{Configuration} \\ [0.5ex] 
 \hline\hline
 BTB Size &  0.5K, 1K, 2K, and {\bf 4K} Entries\\
 Prefetcher & {\bf On}, Off\\
 L2 (private) cache & 256K, 512K, 768K, and {\bf 1024K} Bytes\\ 
 L3 (shared LLC) cache  & 4M, 8M, 12M, and {\bf 16M} Bytes\\
 \hline\hline
\end{tabular}
}
\caption{List of reconfigurable hardware resources. Initial configuration is in bold-face.}
\label{table-knob}
\end{table}

\subsubsection{Selecting Hardware Telemetry}
\label{sec-tele}
Modern processors provide hundreds of hardware event counters as the telemetry. Not all of them are relevant in deciding how to reconfigure various resources. Therefore, to select the most relevant ones, we use Pearson correlation coefficient.
We extract a set of 24 microarchitectural counters closely related to those four hardware resources that we want to optimize. These 24 counters capture both program characteristics and their interaction with system components. 
We calculate correlation coefficient among the counters.
Correlation coefficient gives numerical measure of the effectiveness of the statistical relationship between any two counters. It estimates how strong or weak the relationship between any two counters is. High positive or negative value of correlation indicates high level of positive or negative dependence between two counters.
We use correlation coefficient to select the most informative counters. Section~\ref{sec-feature-select} shows the detailed coefficient results of these counters.



\subsubsection{Building Dataset}
\label{sec-data}
With 4 reconfigurable resources, there are $\mathbb{N}=4*2*4*4=128$ possible configurations. Each application is executed and profiled under each of these configurations. During the execution of an application, \scheme\ collects hardware telemetry, calculates energy efficiency periodically after every $\mathbb{I}$ instructions 
and records them in a profiling file. 
Let us call every $\mathbb{I}$ instruction an {\em Interval}. Let us denote the telemetry as $\mathbb{T}=\{t_i\}^n_{i=1}$, where each $t_i$ is an individual hardware counter and the energy efficiency as $\mathbb{E}$. Thus, the profiling file contains a set of records of  $<\mathbb{T}, \mathbb{E}>$, one record for each interval. 
\scheme\ keeps the input fixed for an application during profiling.
Still, there could be slight perturbation during some execution
due to the difference in hardware configurations and thread scheduling (in case of a multithreaded application).
Therefore, we choose $\mathbb{I}$ to be large enough so that the number of intervals remains the same in every profiling file of an application. As a result, each corresponding interval in different profiling files represents (roughly) the same code region of the application. Whatever little difference that could exist among the code regions of similar intervals, adds noise to the training dataset. Such noise works in favor of DNN models to improve their accuracy.

\begin{figure}[htpb]
\centering
\includegraphics[width=0.6\columnwidth]{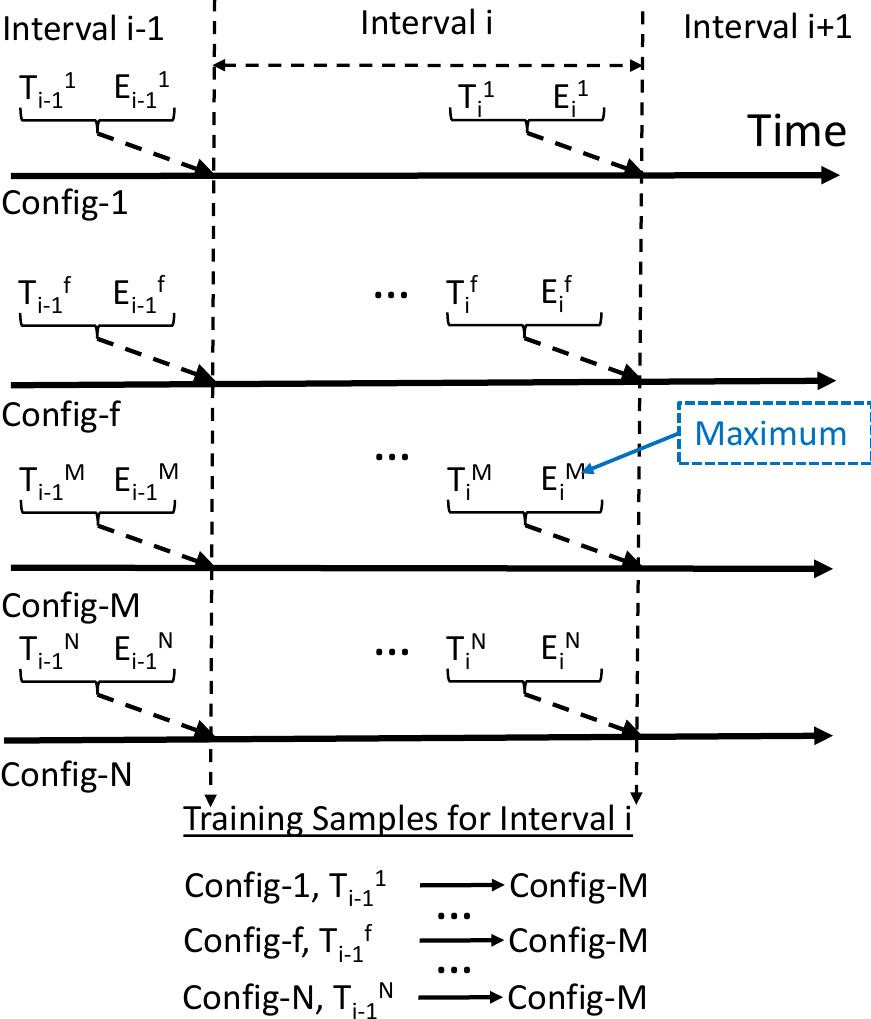}
\caption{How training samples are formed from profiles.}
\vspace{-0.7cm}
\label{fig-input-form} 
\end{figure}

Let us consider an interval $i$. The profiling record for $i$ is $<\mathbb{T}_i^{f}, \mathbb{E}_i^{f}>$ in the 
profiling file for $Config-f$ ($Config-f$ could be any of the $\mathbb{N}$ configurations i.e., $1\le f\le \mathbb{N}$). 
\scheme\ finds the maximum among $\mathbb{E}_i^1$ to $\mathbb{E}_i^{\mathbb{N}}$. The configuration 
corresponding to the maximum, say $Config-\mathbb{M}$, provides the highest energy efficiency. Therefore, at runtime, 
when \scheme\ tries to predict the best configuration at the beginning of interval $i$, it should predict $Config-\mathbb{M}$ 
as the output of the DNN model. That is why, phase 1 forms a training sample by using $\mathbb{T}_{i-1}^f$ as the 
input and $Config-\mathbb{M}$ as the output. Note that \scheme\ uses $\mathbb{T}_{i-1}^f$ instead of $\mathbb{T}_i^f$ as 
the input because the telemetry collected at the beginning of interval $i$ is the telemetry corresponding to interval $i-1$. 
Thus, DNN should be trained to predict $Config-\mathbb{M}$ (the best configuration for interval $i$) by using the telemetry 
collected at the beginning of interval $i$. 
Figure~\ref{fig-input-form} shows the telemetry and energy efficiency of different intervals across different configurations.
Last but not least, in addition to $\mathbb{T}_{i-1}^f$, the corresponding 
configuration i.e., $Config-f$ is provided as part of the input in the training sample. In other words, the training sample is formed 
by using $<Config-f, \mathbb{T}_{i-1}^f>$ as the input and $Config-\mathbb{M}$ as the output. In total, there are $\mathbb{N}$ training
samples for interval $i$.

\begin{figure*}[htpb]
\centering
\includegraphics[width=0.6\linewidth]{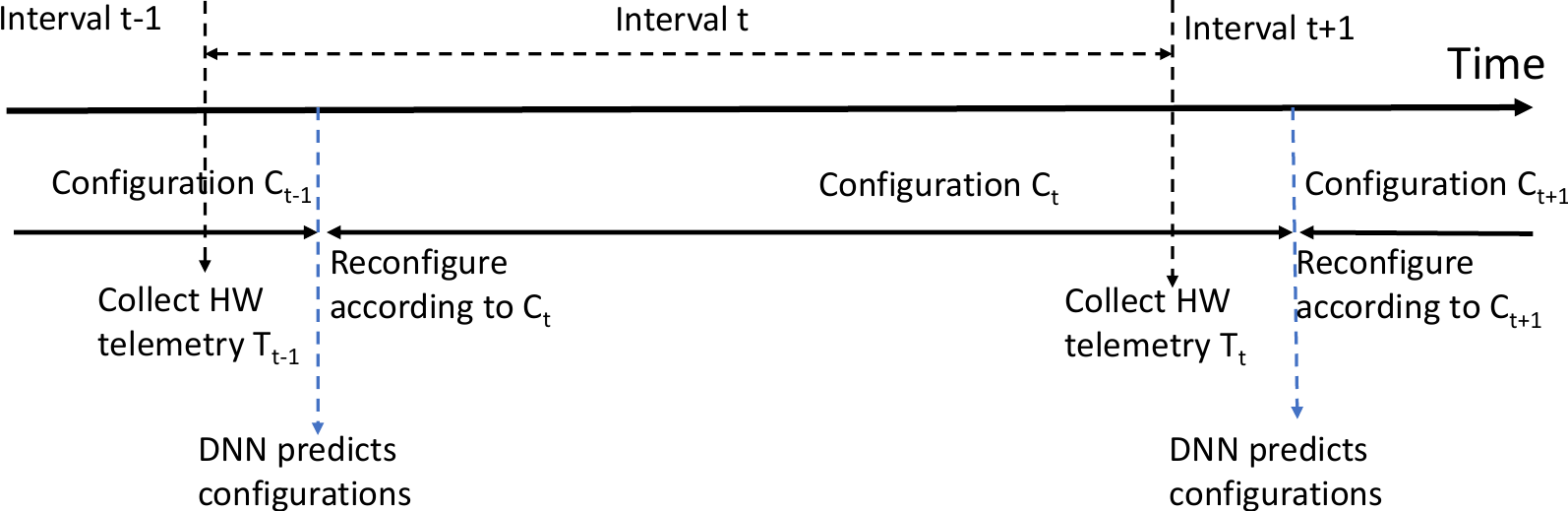}
\caption{Timing of various steps of \scheme.}
\vspace{-0.5cm}
\label{fig-timing} 
\end{figure*}

\subsubsection{Selecting a Predictive Model}
\label{sec-pred-model}
The dataset built in Section~\ref{sec-data} is used to train a machine learning model. We initially experimented with simpler models such as logistic regression or maximum likelihood estimation. Our experiments revealed that such simpler models are not able to improve energy efficiency of the system (Section~\ref{sec-efficiency-eval}). Therefore, we propose to use a DNN model.
In order to find the best model, \scheme\ searches all possible network configurations within a constrained search space (e.g., all topologies up to the maximum of $H$ hidden layers and $L$ neurons per layer) and picks the one with the highest accuracy. 

\subsection{Phase 2: Prediction-based Hardware Reconfiguration}
\label{sec-reconfig}
During this phase, \scheme\ loads the trained DNN model in a DNN hardware and uses it to predict the configuration of hardware resources for maximizing the energy efficiency. When an application starts execution, \scheme\ starts with maximum resources. This prevents any initial slowdown due to insufficient resources. \scheme\ collects hardware telemetry after every interval of $\mathbb{I}$ instructions. Suppose the telemetry after interval $t$ is $\mathbb{T}_t$ and the resource configuration is $C_t$. \scheme\ uses the DNN hardware with $<C_t, \mathbb{T}_t>$ as the input to infer the new configuration $C_{t+1}$. Figure~\ref{fig-timing} shows the timing of inference step. After DNN hardware calculates the predicted configuration, $C_{t+1}$, \scheme\ reconfigures the hardware resources according to $C_{t+1}$. Now, we describe how each resource is reconfigured.

\subsubsection{L2 and L3 Caches}
\label{sec-caches}

Caches, mainly designed in SRAM (we are not considering eDRAM in this paper) are sources of both static and dynamic power consumption.
Dynamic power is consumed in row-decoder, column-decoder, pre-charge circuit and some parts of the core cell and depends on access pattern.
Static power, mainly leakage, is dissipated in every cell of the SRAM cache.
With the continuous reduction in transistor sizes and, consequently, the switching threshold voltage of the transistor, static power becomes the major source of power dissipation in caches~\cite{cache-static}.
Therefore, when we turn-off parts of the cache, we want to ensure that we target leakage current.
For that, we use gated-ground~\cite{drg-cache, dri-cache,cache-vlsi}.

\begin{figure}[htpb]
\begin{center}
\includegraphics[width=\columnwidth]{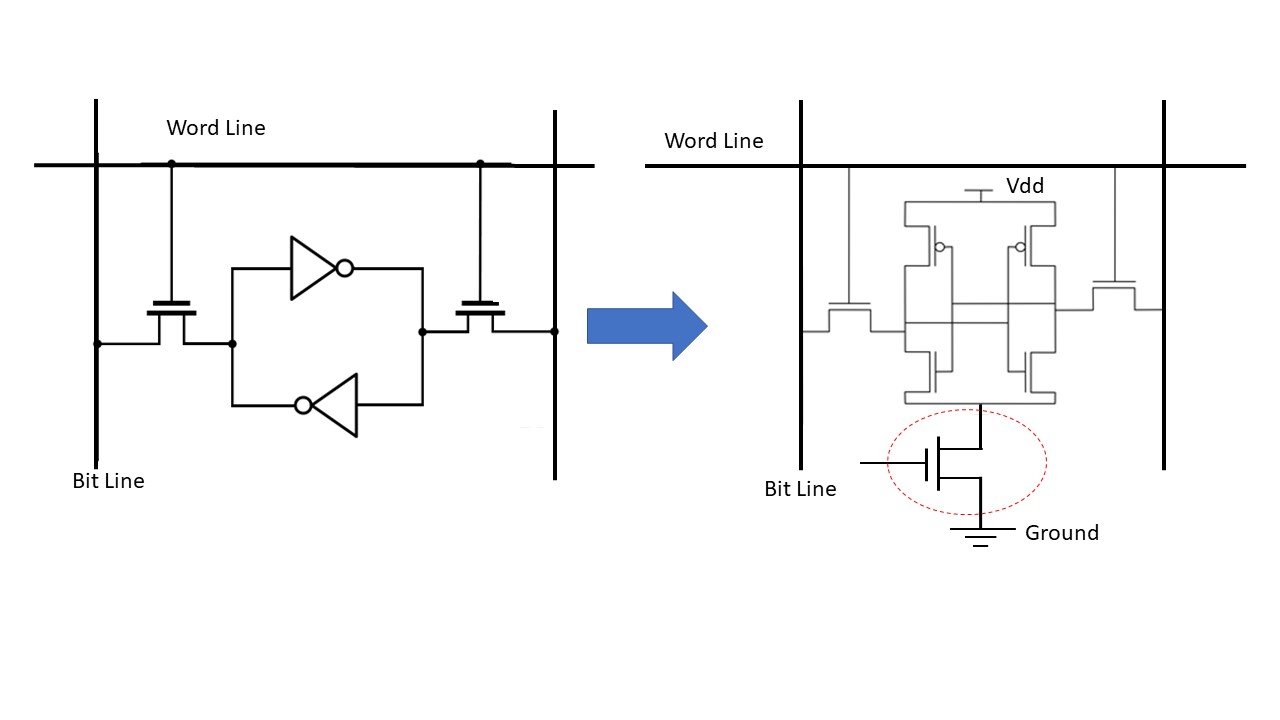}
\vspace{-1.5cm}
\caption{SRAM Cell Design (6T-MC) with the gated-Vdd shown on the Right}
\vspace{-1cm}
\label{fig:sram}
\end{center}
\end{figure}

There are several ways of implementing SRAM cells. 
The one most widely used, due to its relatively high noise immunity, is 6-transistors Memory Cell (6T-MC), shown in Figure~\ref{fig:sram}.
The left part of the figure shows the gate-level of a single SRAM call.
The right part shows the circuit level. 
There is an extra transistor, shown circled, that is used to reduce leakage current that constitutes the major part of the static power dissipation in caches.
In our design of cache resizing, we turn-off individual blocks and never a full-set.
Therefore, we can use a single transistor per block. That is, one extra transistor per 64 cells for a 64-byte block.
This design does not use more than extra 4\% of area overhead with around 5\% increase in cache latency~\cite{drg-cache}.
The increased access latency has been taken into consideration in our simulation.
When a block is turned-off, that extra transistor is also turned off causing a {\em stacking effect} that reduces leakage current by orders of magnitude~\cite{drg-cache}.

\begin{figure}[htpb]
\centering
\includegraphics[width=\columnwidth]{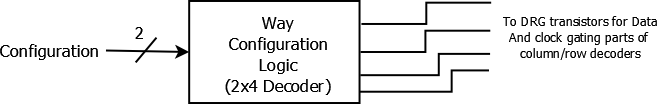}
\caption{Logic for reconfiguring L2 and L3 caches.}
\vspace{-0.6cm}
\label{fig-cache} 
\end{figure}

The next step is to control which blocks will be gated (for static power) and control which parts will be clock-gated to avoid accessing the blocks that are turned-off.
From Table~\ref{table-knob}, we can see that we have four configurations for the cache. 
We need two bits to represent those configurations.
A 2x4 decoder is enough, as shown in Figure~\ref{fig-cache}.
The output of the decoder that is set to one, is used to turn off the corresponding transistors in the data lines.
We turn off blocks starting from the last way in each set.
For example, in LLC cache, if we want to go from 16MB to 12MB in a 16-way, we turn-off ways 15, 14, 13, and 12.
Before this happens, the cache controller checks the dirty bits.
Dirty blocks are written back. 
The output of the decoder is also used to clock-gate parts of the column and row decoders to avoid accessing the parts of the cache that are turned off.
The reconfiguration of the cache does not happen in the critical path of the execution. 
Therefore, it does not have an effect on performance, except the negligible area and latency increase stated above.

\subsubsection{Branch Target Buffer (BTB)}
\label{sec-btb}
\begin{figure}[htpb]
\centering
\includegraphics[width=0.5\columnwidth]{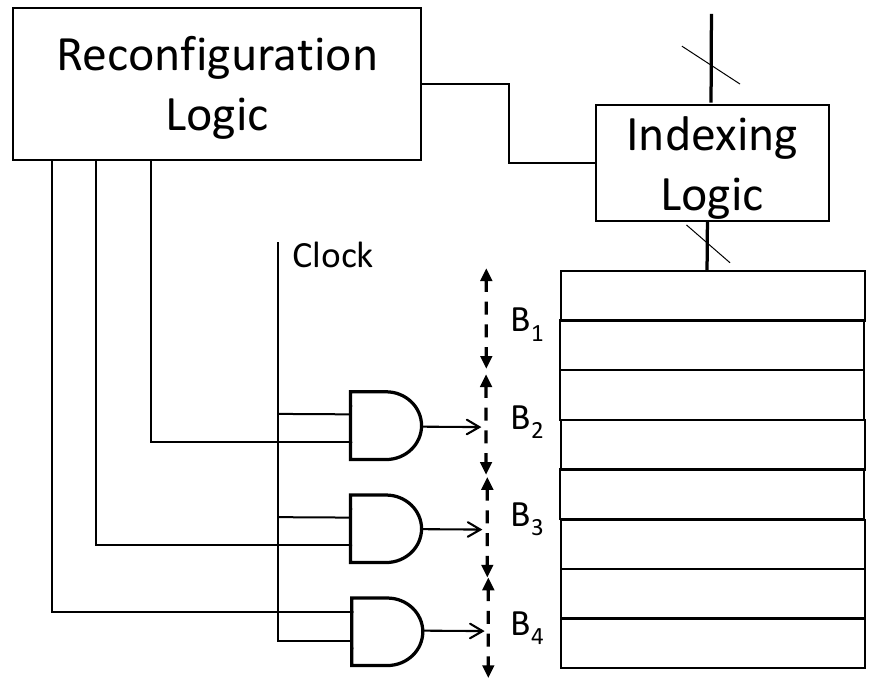}
\caption{Logic for reconfiguring the BTB.}
\vspace{-0.2cm}
\label{fig-BTB} 
\end{figure}

BTB has 4 possible configurations (Table~\ref{table-knob}). Therefore, we can partition BTB into 4 sections - $B1$, $B2$, $B3$, and $B4$ (Figure~\ref{fig-BTB}). For the first configuration (i.e., 0.5K entries), sections ($B2$, $B3$, $B4$) are clock-gated. Similarly, for the second and third configurations, sections ($B3$, $B4$) and ($B4$) are clock-gated respectively. The last configuration does not clock-gate any section at all. On the other hand, Section $B1$ is never clock-gated because at least those entries in BTB are used in all configurations. We add a reconfiguration logic that creates the appropriate clock-gating signal to enable the appropriate sections. Moreover, for each configuration, the indexing logic needs to reconfigure the indexing bits accordingly. The extra logic circuits add negligible latency. In a multicore processor with one BTB per core, \scheme\ reconfigures all BTBs to the same configuration. This is done is to simplify the prediction and reconfiguration logic in \scheme.

\subsubsection{Prefetcher}
\label{sec-prefetcher}
Prefetcher is used either completely or not at all. Therefore, the prefetcher is clock-gated entirely or not at all. So, the reconfiguration logic simply generates a single clock-gating signal for the entire prefetcher.

\section{Implementation}
\label{sec:impl}
In this section, we outline the implementation of DNN in \scheme. 
Recent trends in architecture suggest that upcoming machines will have one or more DNN accelerators in the 
form of FPGA or ASIC. That is why we propose an FPGA-based design. We implement a simple DNN accelerator in FPGA and
add a CPU-side DNN Driver Module to control the operation of the accelerator. The module forms inputs,
collects outputs from the accelerator, and sends control signals to the FPGA.

\subsection{DNN Accelerator}
\label{sec-dnn}
\begin{figure}[htpb]
\centering
\includegraphics[width=0.9\columnwidth]{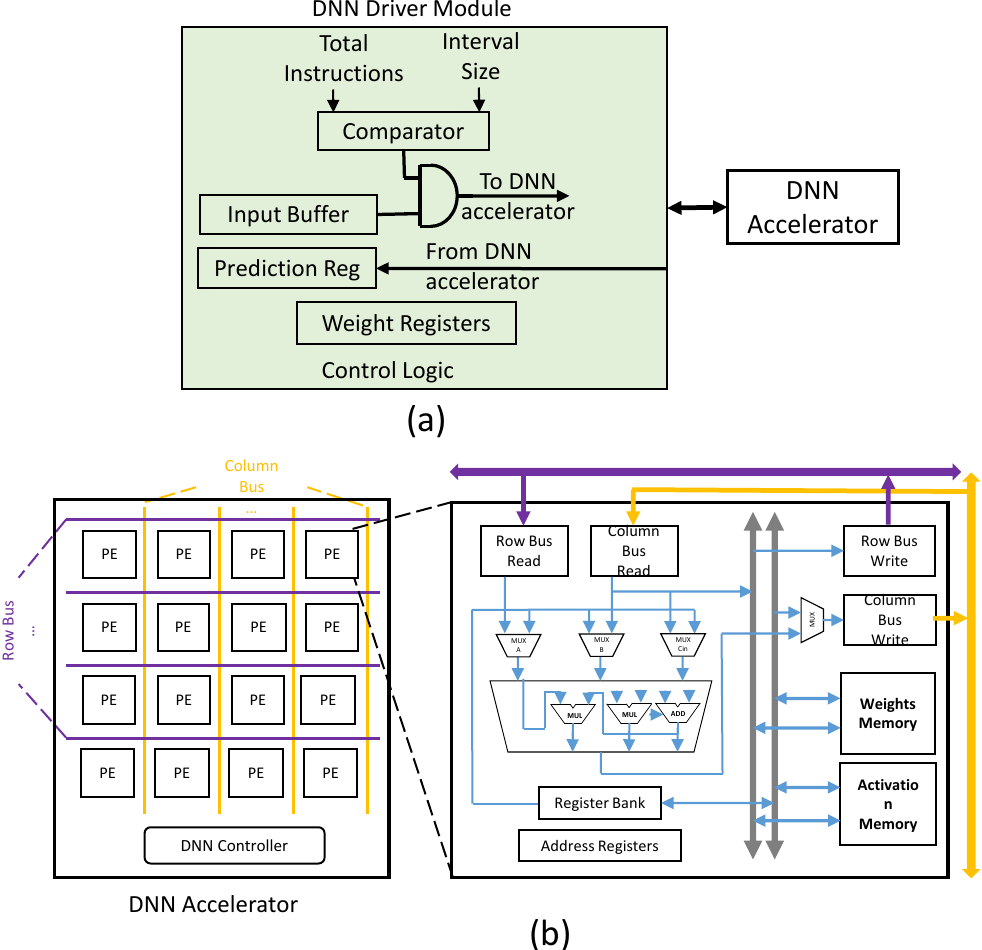}
\caption{Details of the DNN module.}
\vspace{-0.3cm}
\label{fig-dqn} 
\end{figure}

There are many DNN accelerator designs in literature ~\cite{hadi12, eyeriss, Reagen2016}. We used one
similar to the one proposed by Yuanfang Li~\cite{caterpiller}. Figure~\ref{fig-dqn} shows the overview of our design. 
The accelerator is constructed as a systolic array of Processing Elements (PEs). The systolic array supports the fast
broadcast of inputs and partial sum generation using row and column buses.
Each PE contains 2 memories for storing activations and weights, 2 Multiplier, 1 Adder,
2 output buffers for sending results to the row and column buses, 2 input buffers for loading data from row and column buses, 
and 4 multiplexers for handling reduction operation during the forward  propagation.
We consider an extra module for calculating the Softmax function.
We distribute the weights of all layers among PEs. 


Inference operation is done in 5 steps. {\em Step 1,} the DNN Controller broadcasts the 
input vector across the row buses and calculates the related address of weight memory 
in each PE to compute matrix-vector multiplication. 
{\em Step 2,} 
each PE performs a multiplication operation and adds any prior result from the same PE, and then, loads the next part of the input and related weights from memory. 
{\em Step 3,} Steps 1-3 are repeated until the input vector is completely processed. {\em Step 4,} each PE broadcasts partial-sums along the column bus to perform a column-wise reduction.
{\em Step 5,} using a multiplexer, the accelerator applies RELU activation function on outputs and then, stores the outputs in activation memory inside the PEs. We repeat steps 1-5 until all the results become ready. These activations are inputs for the next layer. We repeat this process for next layers until the last layer. 

\subsection{DNN Operation}
\label{sec-dqn}
The DNN driver module has an input buffer, a prediction register, weight registers, parameter registers, and a control logic. 
The input buffer is responsible to generate inputs that are provided to the DNN accelerator to infer the predictions. 
An input consist of various hardware counters and current configuration.
Each core collects the counters and current configuration of resources independently and sends them to the driver module after every $n$ (e.g., say $n$=10,000) instructions. When the module receives counters of at least a total of $\mathbb{I}$ (e.g., $\mathbb{I}$ = interval size) 
instructions, \scheme\ assumes the start of a new interval. The module aggregates the counters and normalizes each counter with respect to the total instructions of the interval that just finished.
The driver module then sends the formed input to the DNN accelerator and receives the predicted configuration. The control logic sends the new configuration to the cores and cache controllers to initiate the reconfiguration.


The control logic also contains a set of registers - weight registers and prediction register. Weight registers can be read and written using load and store instructions to initialize the DNN accelerator with a certain trained model. The prediction register stores the predicted configuration returned by the DNN accelerator. 

\section{Experimental Setup}
\label{sec:eval}
We used Multi2Sim~\cite{multi2sim} and McPAT~\cite{mcpat} to simulate the experimental hardware and its power consumption. We implemented DNN hardware in Xilinx FPGA to calculate the latency and overhead of DNN hardware. This latency is used inside Multi2Sim.
Table~\ref{table-params} shows the parameters of the simulated hardware that we use to conduct the experiments. 
We used PARSEC 3.0 benchmarks with small inputs to generate training data and evaluate the performance of our approach. 
All benchmarks are run to completion or 1.0 billion instructions. The interval size $\mathbb{I}$ is set to 0.5 million instructions. 


\begin{table}[h!]
\centering
\begin{tabular}{||l| l ||} 
 \hline\hline
 Parameter & Value \\ [0.5ex] 
 \hline\hline
 CPU & 8-core @ 2.4Ghz, SMT off\\ 
 Private L1 cache (I/D) & 32KB, 64B line, 8-way\\
 Private L2 Cache       & 1024K, 64B line, 16-way\\
 Shared L3 Cache        & 16M, 64B line, 16-way\\
 Coherence Protocol     & Directory-based MOESI\\[1ex]
 \hline\hline
\end{tabular}
\caption{Parameters of the simulated hardware.}
\label{table-params}
\end{table}

We experimented with both single and multiprogram scenarios to evaluate \scheme. In case of the single program scenario, we used 8 Parsec applications. We used leave-one-out cross-validation approach. For the multiprogram scenario, we evenly split the Parsec applications into two sets - one set for training and the other for testing. We randomly selected 5 combinations of programs. Each combination contains 4 different programs. Two instances of each program are launched during the execution of that combination. Table~\ref{tbl:comb} shows the combinations.

\begin{table}[h!]
\centering
\scalebox{0.9}{
\begin{tabular}{||c|l||} 
 \hline\hline
 Combination & \multicolumn{1}{c||}{Programs} \\  
 \hline\hline
1 & bodytrack, facesim, freqmine, swaptions\\\hline
2 & canneal, facesim, streamcluster, swaptions \\\hline
3 & bodytrack, canneal, freqmine, streamcluster \\\hline
4 & facesim, canneal, freqmine, swaptions \\\hline
5 & bodytrack, canneal, facesim, swaptions \\\hline
 \hline
\end{tabular}
}
\caption{Multiprogram combinations used in evaluation.}
\label{tbl:comb}
\end{table}


For comparison purpose, we implemented the Maximum Likelihood Estimation (MLE) model from~\cite{dubach10}. 
We call this \textit{MLE-original}. Additionally, We implemented another version of \scheme\ using an MLE model instead of a DNN model. This is to compare the performance of the DNN model to the simpler MLE model. We call it \textit{MLE}. We also compared our scheme with the DVFS algorithm from~\cite{dvfs}.

\section{Results}
\label{sec:res}
\subsection{Efficiency Evaluation}
\label{sec-efficiency-eval}

\begin{figure}[h]
\centering
\begin{tabular}{c}
    \includegraphics[width=0.8\columnwidth]{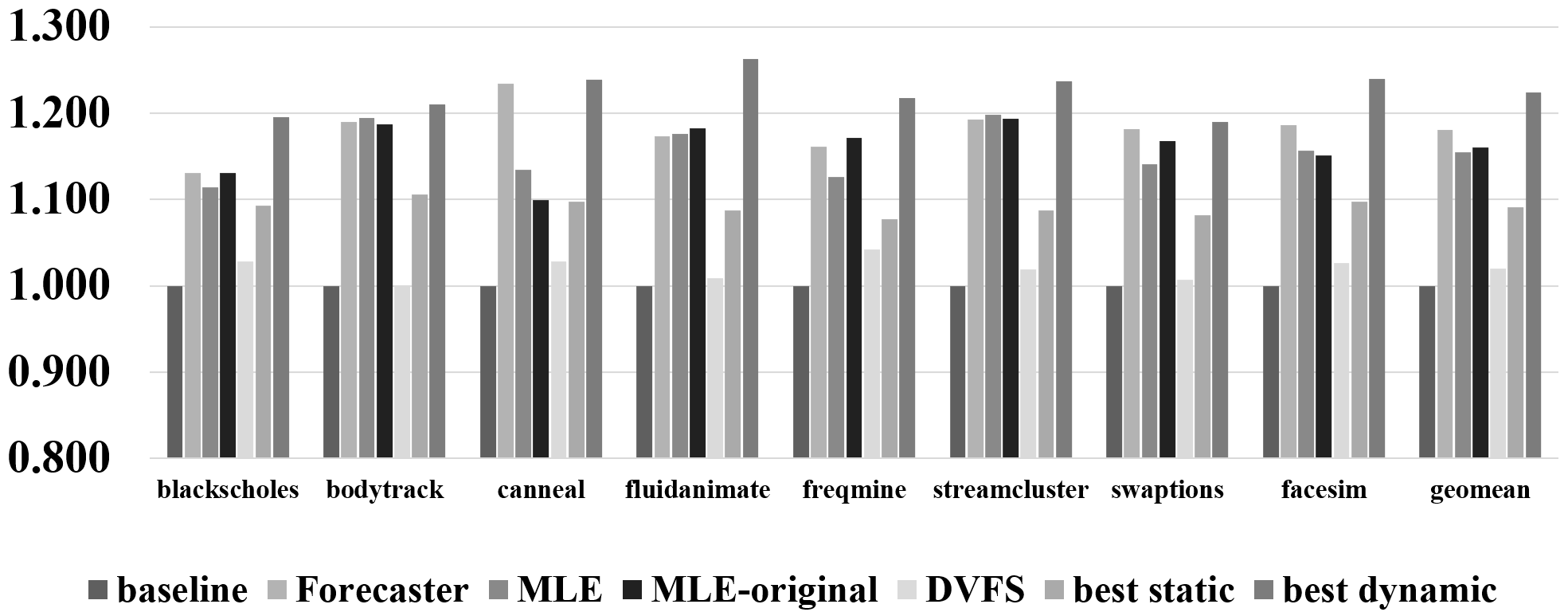}\label{fig-efficiency-single-program}\\
     (a) \\
    \includegraphics[width=0.8\columnwidth]{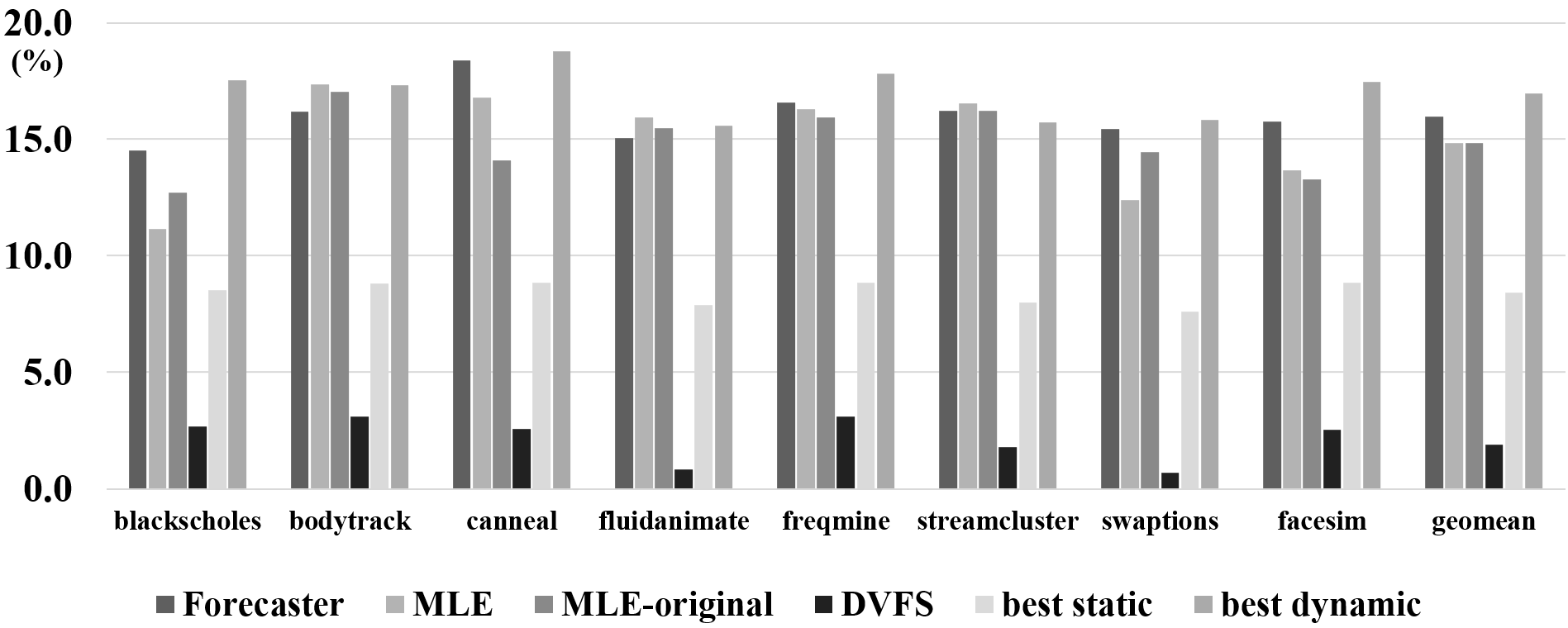}\label{fig-power-single-program}\\
    (b)\\
    \includegraphics[width=0.8\columnwidth]{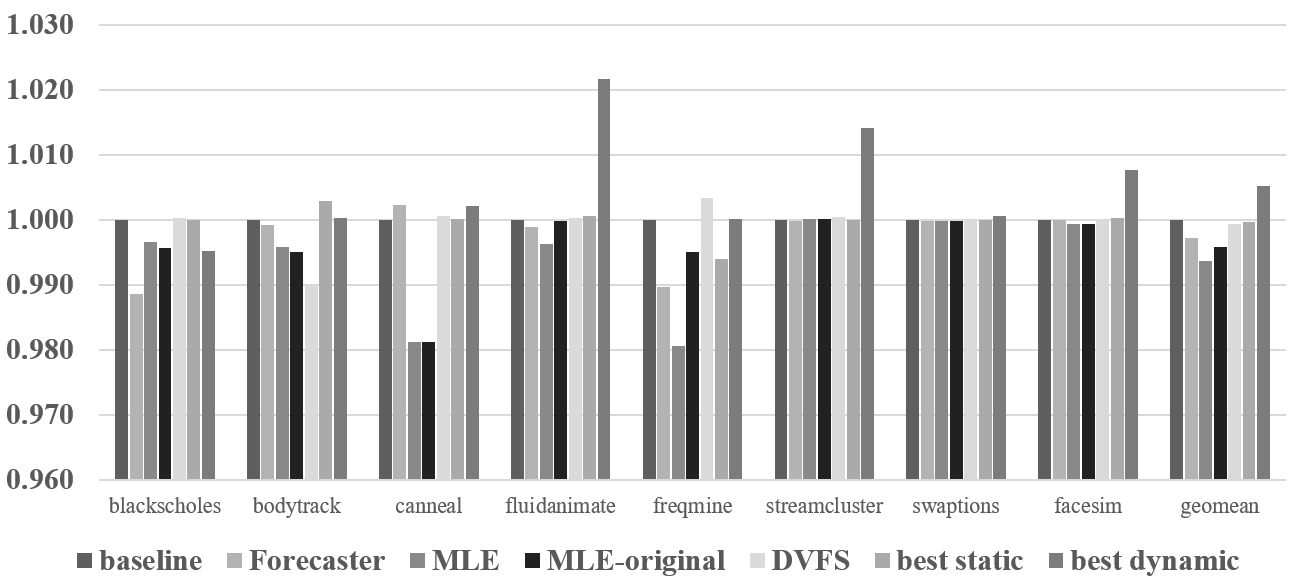}\label{fig-ips-single-program}\\
    (c)\\
\end{tabular}
\caption{ Normalized (a) efficiencies, (b) power, and (c) IPS of different optimization schemes in single program mode.}
\vspace{-0.5cm}
\label{fig-single-program} 
\end{figure}

\begin{figure}[h]
\centering
\begin{tabular}{c}
    \includegraphics[width=0.8\columnwidth]{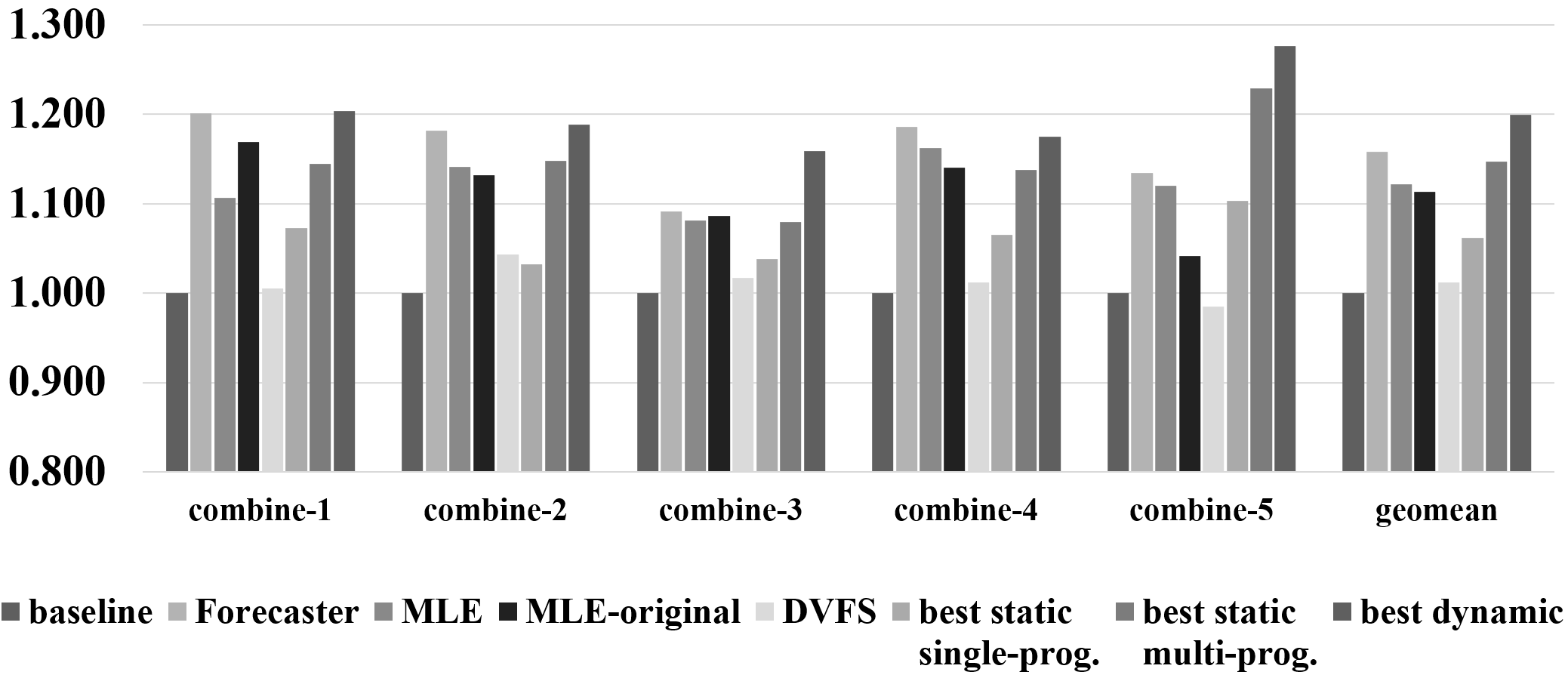}\label{fig-efficiency-multi-program}\\
     (a) \\
    \includegraphics[width=0.8\columnwidth]{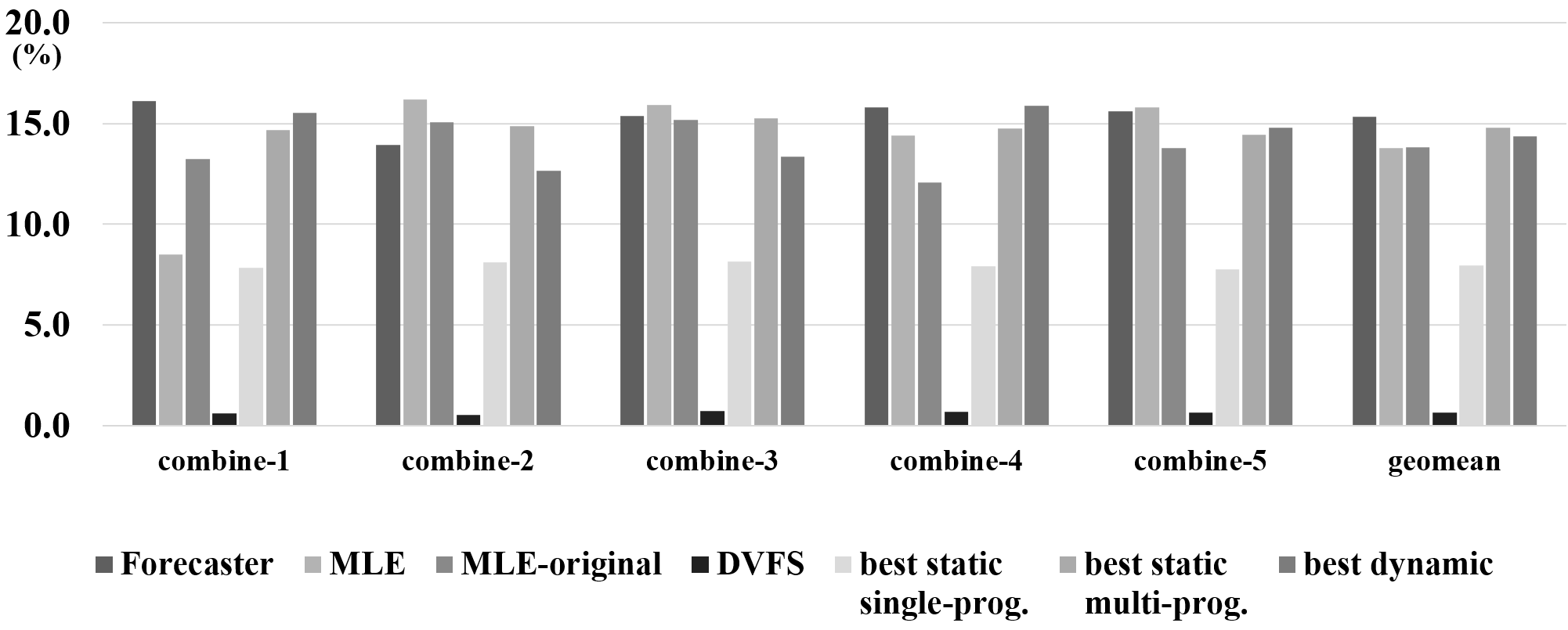}\label{fig-power-multi-program}\\
    (b)\\
    \includegraphics[width=0.8\columnwidth]{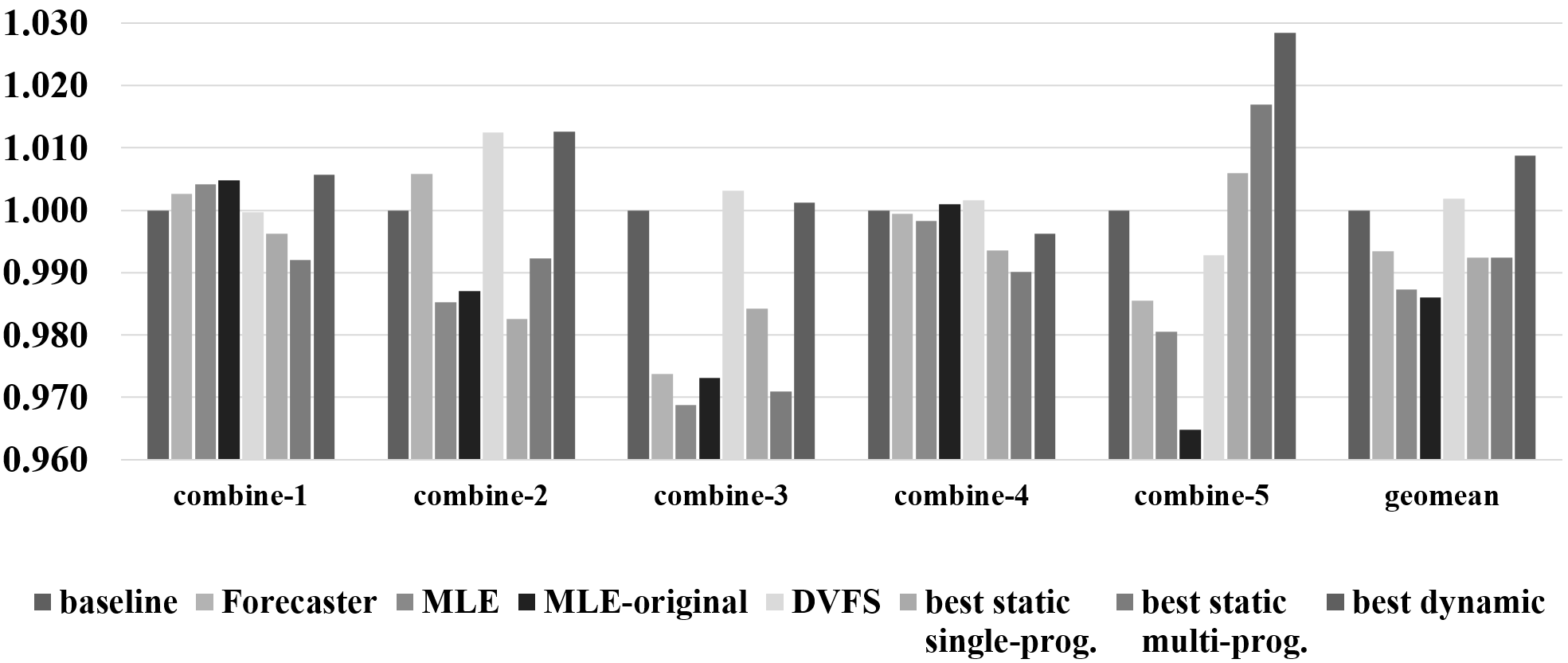}\label{fig-ips-multi-program}\\
    (c)\\
\end{tabular}
\caption{ Normalized (a) efficiencies, (b) power, and (c) IPS of different optimization schemes in multi program mode.}
\vspace{-0.5cm}
\label{fig-multi-program} 
\end{figure}

\begin{figure}[h]
\centering
\begin{tabular}{c}
    \includegraphics[width=0.7\columnwidth]{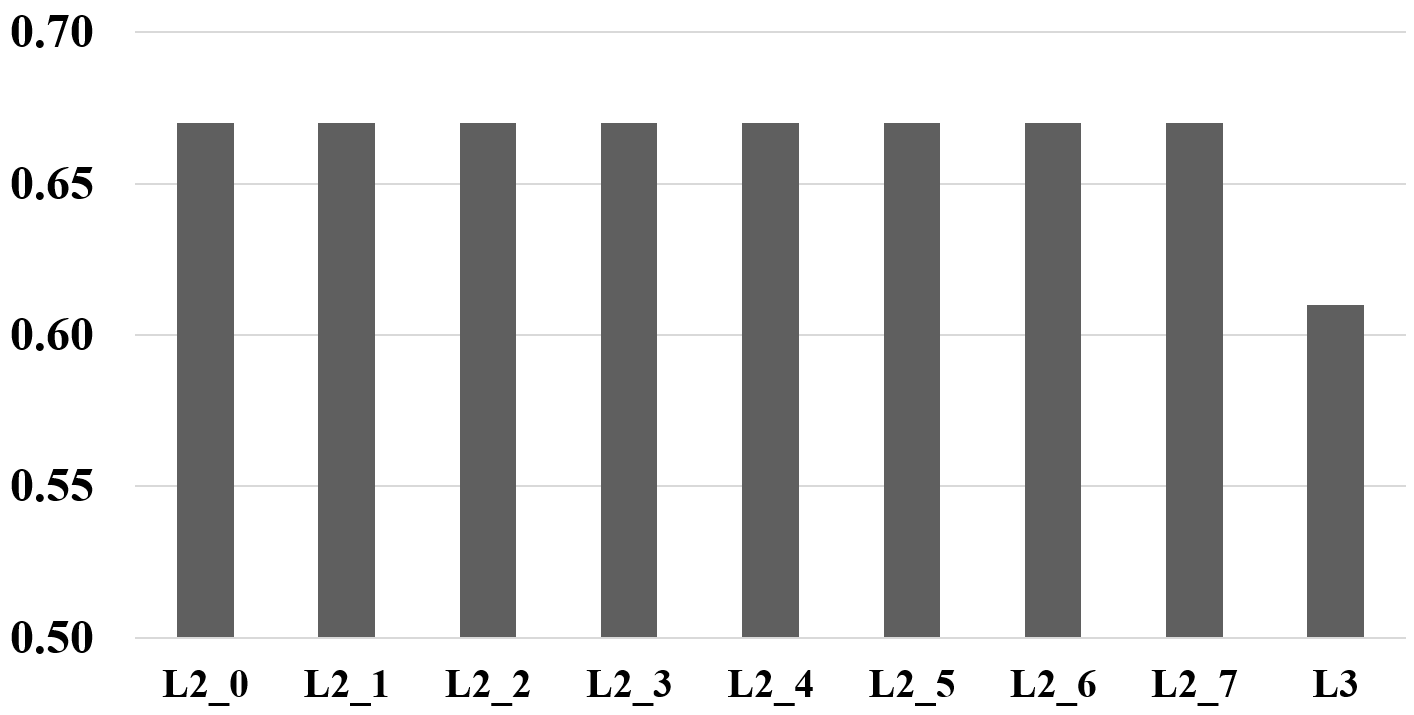}\label{fig-cache-power-single-program}\\
     (a) \\
    \includegraphics[width=0.7\columnwidth]{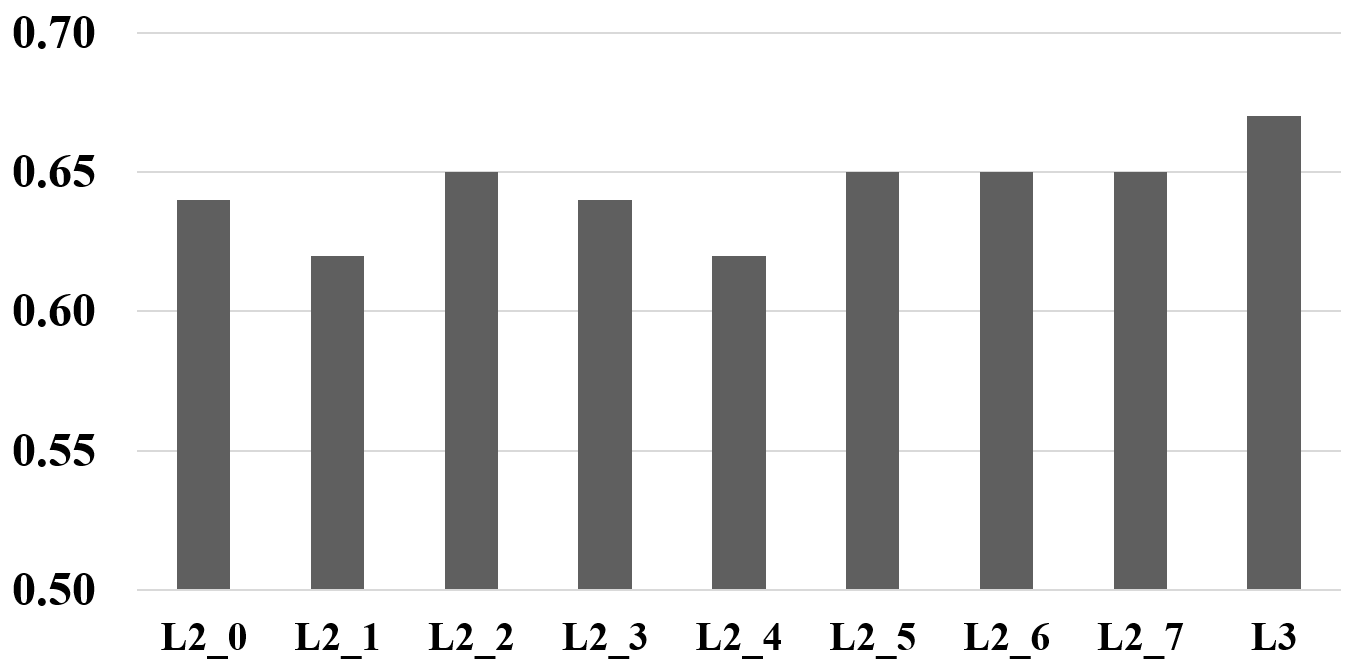}\label{fig-cache-power-multi-program}\\
    (b)\\
\end{tabular}
\caption{Average percentage saving in cache static power of (a) \textit{swaptions} (single-program), (b) \textit{combine-5} (multi-program)}
\vspace{-0.6cm}
\label{fig-cache-power-saving} 
\end{figure}

The efficiencies of single program and multi program experiments are shown in Figure~\ref{fig-single-program}(a) and Figure~\ref{fig-multi-program}(a), respectively. In both cases, our scheme outperforms all other tuning techniques as well as the best static configuration. On average, \scheme\ improve the system efficiency by 18.1\% compared to the baseline in single program workload and 15.8\% in multiple program workload. These improvements represent 80\% of those of the best dynamic scenario, in which we know ahead of time the optimal setting for all hardware components for each execution phase. 

These efficiency gains can only be achieved thanks to the capability of \scheme\ to accurately predict the hardware demand of applications in each phase. As a result, it can save the most possible amount of power while yielding the same performance as the baseline, as shown in Figure~\ref{fig-single-program}(b)(c) and Figure~\ref{fig-multi-program}(b)(c). The performance degradation of \scheme\ is only 0.003\% in single program mode and 0.007\% in multi program mode. Meanwhile, \scheme\ manages to save 16\% and 15.3\% in power compared to the baseline in single and multi program scenarios, respectively. This power savings is 6\% and 7\% more than the MLE and MLE-original techniques respectively. For multi program workload, \scheme\ even preserves more power than the best dynamic configuration, which only manages to save 14.3\%.

\subsection{Detailed Analysis}

\begin{figure}[h]
\centering
\begin{tabular}{c}
    \includegraphics[width=0.7\columnwidth]{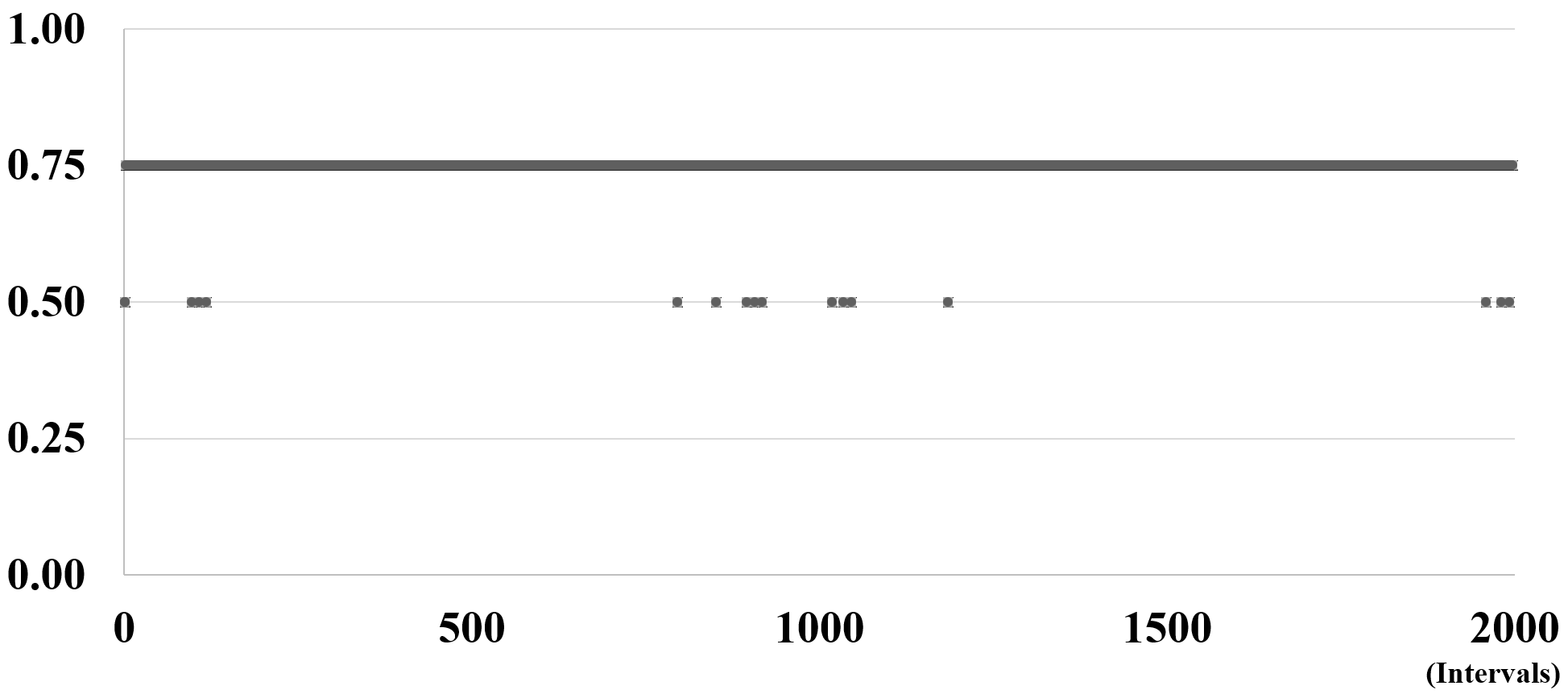}\label{fig-swap-L2}\\
     (a) \\
    \includegraphics[width=0.7\columnwidth]{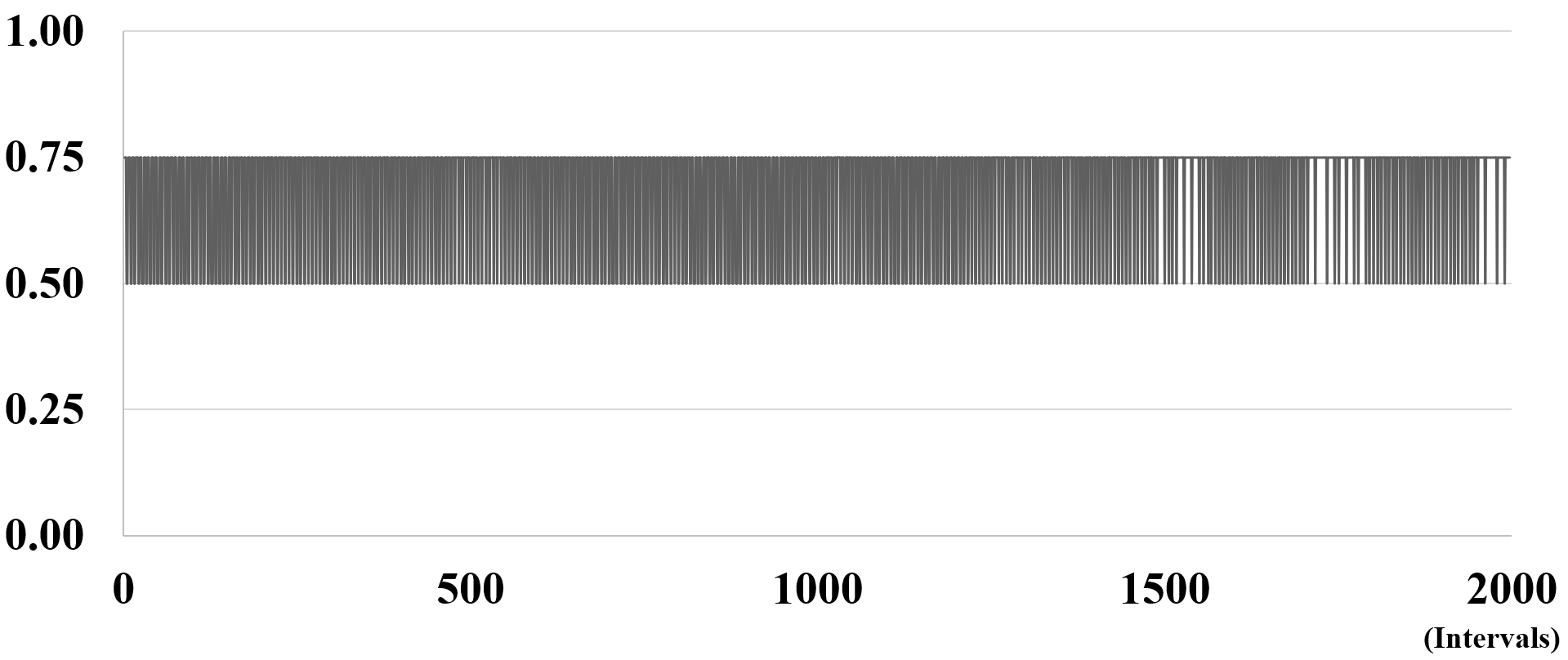}\label{fig-swap-L3}\\
    (b)\\
    \includegraphics[width=0.7\columnwidth]{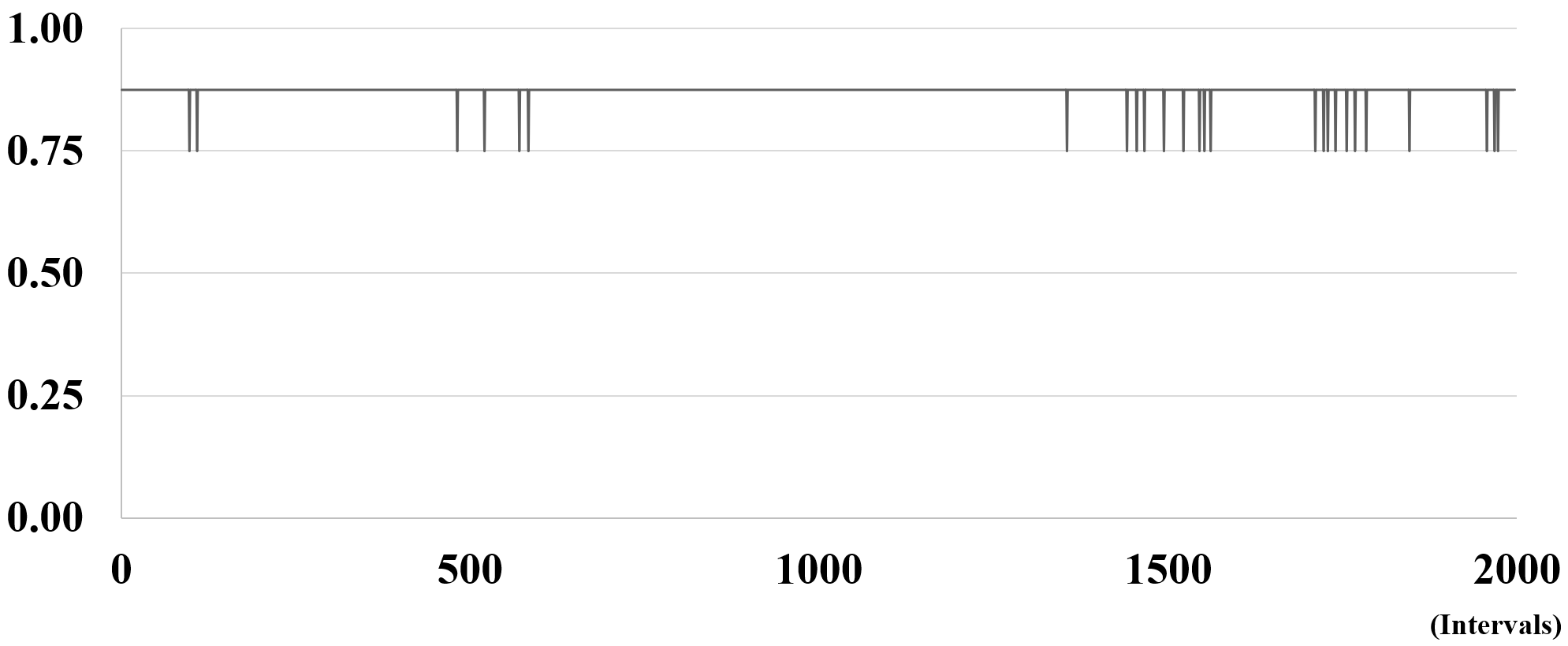}\label{fig-swap-btb}\\
    (c)\\
\end{tabular}
\caption{Avg amount of (a) L2, (b) L3, and (c) BTB turned off during the execution of \textbf{\textit{swaptions}}.}
\vspace{-0.5cm}
\label{fig-swaptions} 
\end{figure}

\begin{figure}[h]
\centering
\begin{tabular}{c}
    \includegraphics[width=0.7\columnwidth]{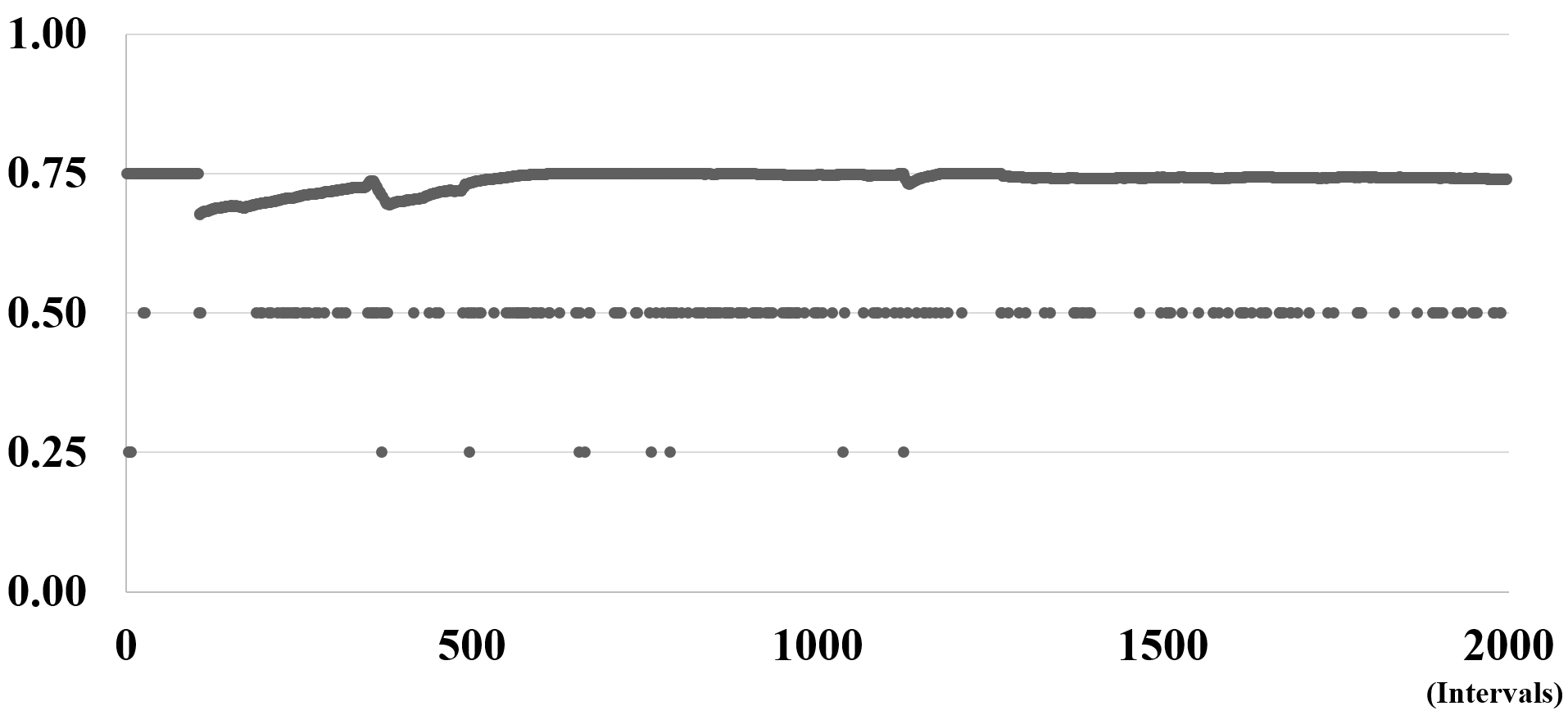}\label{fig-combine5-L2}\\
     (a) \\
    \includegraphics[width=0.7\columnwidth]{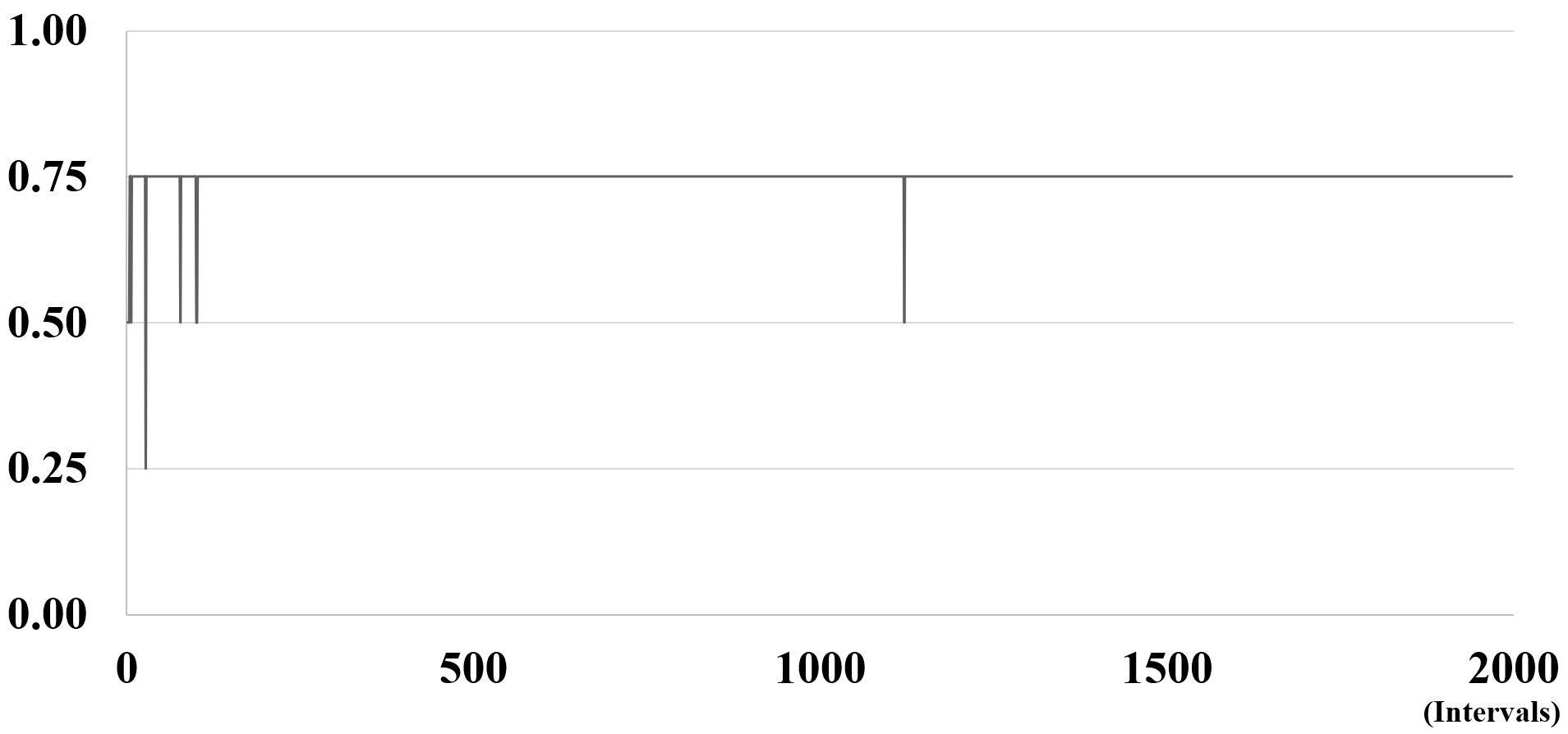}\label{fig-combine5-L3}\\
    (b)\\
    \includegraphics[width=0.7\columnwidth]{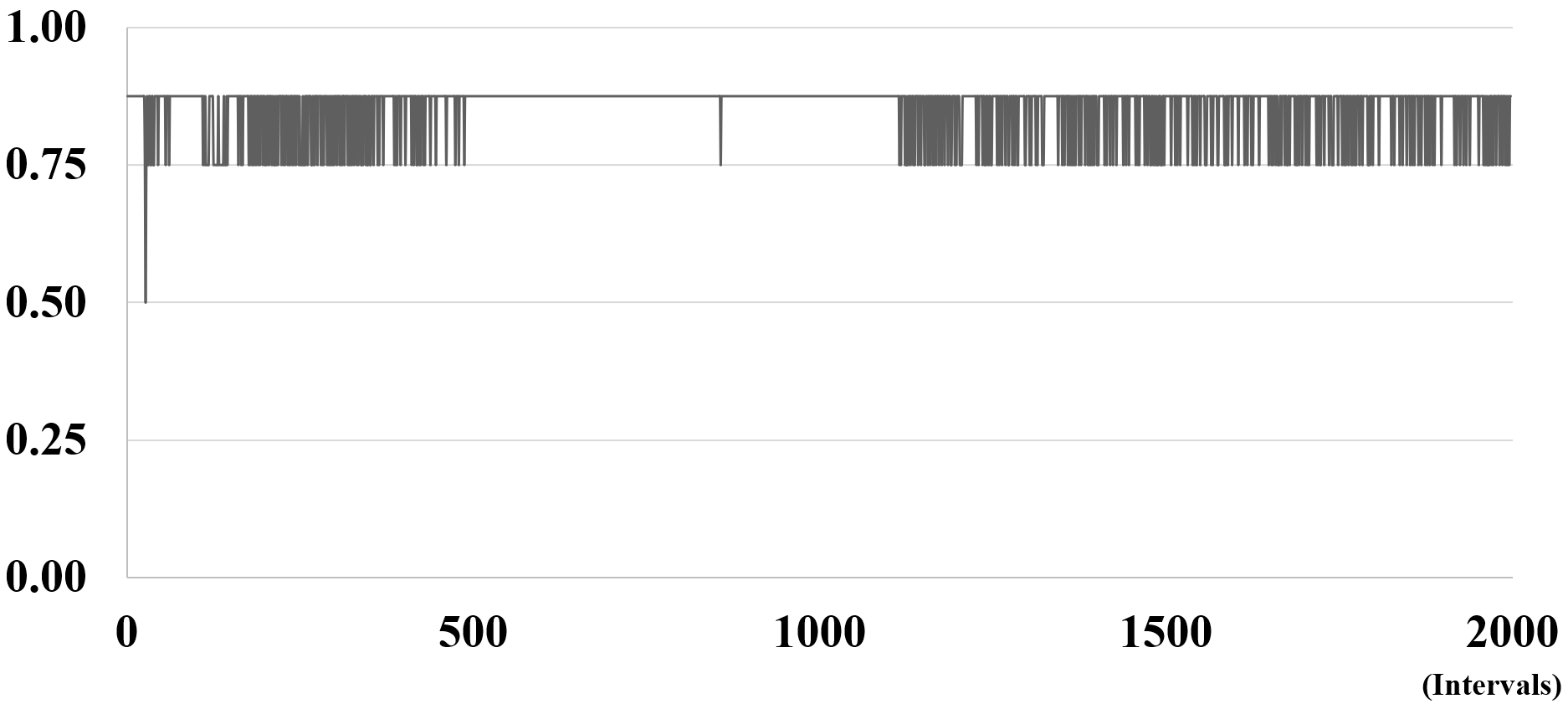}\label{fig-combine5-btb}\\
    (c)\\
\end{tabular}
\caption{Avg amount of (a) L2, (b) L3, and (c) BTB turned off during the execution of \textbf{\textit{combination-5}}.}
\vspace{-0.5cm}
\label{fig-combine5} 
\end{figure}

Figures~\ref{fig-swaptions} and~\ref{fig-combine5} show how \scheme\ manages the hardware resources during program execution in single (\textit{swaptions}) and multi program (\textit{combination-5}) scenarios. In \textit{swaptions}, \scheme\ can turn off 75\%, 68\%, 87\% of L2 cache, L3 cache and the BTB respectively. \scheme\ also deactivates the L1 data prefetcher for 26\% of all intervals. Similar behavior can be seen in multi program scenario, in which 71\% of L2, 75\% of L3 and 85\% of BTB can be disabled without affecting the system IPS. Since we use small input, \textit{swaptions} does not need the full capacity of either caches. \scheme\ accurately estimates the demand of \textit{swaptions}, then turns off as much redundant cache memory as possible, thus save a lot of power while retaining the same performance. Sometimes \scheme\ decision cannot be fully satisfied as shown in Figure~\ref{fig-combine5}(a). In some intervals, only around 65\% to 70\% amount of L2 cache is disabled even though \scheme\ prediction is 75\%. This is because those cache blocks are active. To preserve performance, we do not forcefully turn off resources which are being used, such as valid cache blocks.
Figure~\ref{fig-cache-power-saving} shows the break down in cache static power savings of \scheme. For single program scenario, the amount of power saved is similar between L2 private caches, since program's workload is uniformly distributed across all cores. On the contrary, the amount of power saved in multi program scenario is different between each L2s because it depends on the application running on the core. In general, using gated-ground technique~\cite{drg-cache, dri-cache,cache-vlsi} to turn off cache blocks , we manage to save approximately 90\% of static power of L2 and L3 caches. In swaptions, since \scheme\ turns off 75\% of L2 and 68\% of L3, the actual amounts of static power saved are 67\% and 61\%, respectively. 

\subsection{Sensitivity Analysis}

\subsubsection{Feature Selection}
\label{sec-feature-select}
We compute the absolute value of Pearson correlation coefficient between the input features and the output label, which is \textit{efficiency}. After doing some experiments and analyzing the results we decided to select features having absolute correlation coefficient value greater than 0.20. Table~\ref{table-feature-correlation} shows the features that have their correlation coefficients greater than the cutoff value. The rest of the features can be discarded as it prevents classifiers from learning redundant information which sometimes might result in classifier giving wrong predictions. Selected smaller set of features decreases computational cost and time since we need to train classifiers on large dataset. Those counters, combined with the last interval configuration which consolidated into 4 inputs, form the final set of 14 input features of our DNN model. 

\begin{table}[h!]
\centering
\begin{tabular}{||l| c ||} 
 \hline\hline
 Features & Correlation Coefficient \\ [0.5ex] 
 \hline\hline
 L2 most usage           & 0.633786\\
 normalized commit float & 0.615692\\
 normalized commit mem   & 0.505695\\
 normalized commit int   & 0.450328\\
 L1 data access          & 0.443464\\
 normalized commit ctrl  & 0.436718\\
 L2 avg eviction rate    & 0.337137\\
 L2 most hit rate        & 0.295210\\
 L3 usage                & 0.264086\\
 branch mispred rate     & 0.242080\\[1ex]
 \hline\hline
\end{tabular}
\caption{Pearson correlation coefficients of selected counters.}
\label{table-feature-correlation}
\end{table}

\subsubsection{DNN Architecture}
We use exhaustive search to find the optimal configuration for the DNN. Tested configurations are shown in Table~\ref{table-dnn-sensitivity}. For each configuration, we use 80\% of the dataset for training, and 20\% for validation. The optimal DNN model is selected based on its validation accuracy. If there is more than one model that have the same accuracy, the smaller network will be chosen. The best model that we found has 4 hidden layers with configuration \textit{384/384/256/256}. The validation accuracy of this network is 74\%.

\begin{table}[h!]
\centering
\begin{tabular}{||l| c ||} 
 \hline\hline
 Parameter & Value \\  
 \hline\hline
 Hidden layers & 1, 2, 3, 4, 5\\ 
 Neurons per layer & 64, 128, 192, 256, 320, 384, 448, 512\\
 \hline\hline
\end{tabular}
\caption{Range of DNN architecture searched.}
\label{table-dnn-sensitivity}
\end{table}

\subsubsection{Interval Size}
Interval size is one of the most important parameter in dynamic hardware optimization. If the interval size is too big, we will miss a lot of hardware tuning opportunities. On the other hand, small interval size may add an excessive number of reconfigurations, thus increases the overall system overhead. As such, determining the optimal interval size is crucial. We test with interval sizes of 0.1M, 0.5M, 1.0M, 5.0M, 10M instructions. Figure {fig-int-size} shows the efficiency of different interval sizes across applications, both single and multi program scenarios. On average, an interval size of 0.5M instructions is the optimal setting, outperforming the second best by 1.4\%. 
\begin{figure}[!h]
\centering
\includegraphics[width=\columnwidth]{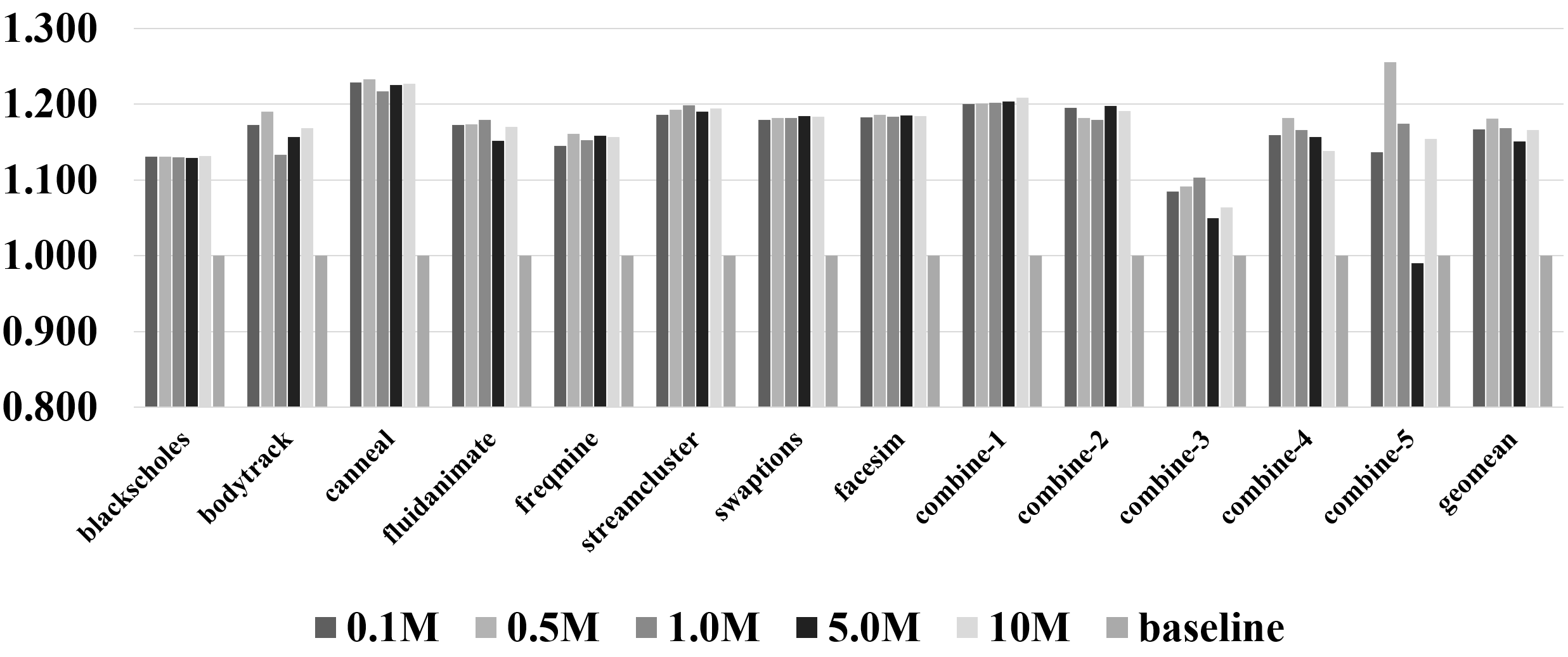}
\caption{Efficiency comparison between different interval sizes}
\label{fig-int-size} 
\end{figure}

\subsection{Overhead Analysis}
The cost of the proposed design can be divided into three parts: delay or latency cost, hardware cost, and power consumption cost.
As for the latency cost, reading the hardware telemetry and making a reconfiguration decision does not happen in the critical path. 
The hardware will continue in its old configuration till the decision is made for a new configuration.

The hardware cost consists of the DNN hardware and the extra hardware used to implement the knobs.
The DNN uses a four-hidden-layer neural network with neuron configuration \textit{384/384/256/256}. 
There is also an input layer of 14 neurons and an output layer of 128 neurons. We considered several design points in terms of the systolic array size when implementing the DNN hardware. Table~\ref{tbl-dnn-over} shows the latency and power consumption of different choices. We used 16*16 PE array size as our implementation choice. Since we envision the DNN hardware to be used in as a general accelerator in many applications, we did not use its overhead in efficiency calculation.

\begin{table}[htpb]
\centering
\scalebox{0.9}{
\begin{tabular}{||c|c|c|c|c|c||} 
 \hline\hline
 \multirow{2}{*}{PE Array} & Frequency & \multirow{2}{*}{Latency} & Slice & \multicolumn{2}{c||}{Power}\\\cline{5-6} 
 & (MHz) & & Reg & Static & Dynamic \\
 \hline\hline
8*8 & 268 & 6352 & 79566 & 196 & 2511 \\\hline
12*12 & 258 & 3200 & 93179 & 211 & 4355 \\\hline
\textbf{16*16} & 247 & 1896 & 109972 & 215 & 4716 \\\hline
 \hline
\end{tabular}
}
\caption{Comparison of different DNN hardware.}
\label{tbl-dnn-over}
\end{table}


The hardware needed for the knobs is straightforward. 
The prefetcher is just clock-gated as the knob is on/off.
The BTB  also uses clock-gating depending on the configuration. 
We have four configurations so a small 2x4 decoder will do the job as shown in the reconfiguration logic of  Figure~\ref{fig-BTB}.
Clock gating the cache ways is simplified by the fact that the way-reconfiguration logic, shown in Figure~\ref{fig-cache}, never gates a valid entry so no change to the cache controller or coherence hardware.
The way-reconfiguration logic is not complicated because it exploits the fact that large caches (such as L3) is usually partitioned.
Therefore we have one logic circuitry per partition.
The power consumption of the above hardware is not high due to several factors.
First, that extra  hardware is activated only at the end of each program phase to make prediction and reconfigure the knobs.
Second, the extra power consumption is much smaller than the power-saved by gating the reconfigured structures.


\section{Conclusions}
\label{sec:conc}
This work presents the potential of dynamically tuning hardware components to save power with negligible performance overhead. With a well-trained deep neural network, our scheme, \scheme\, can save up to 18.4\% of power consumption compared to the baseline configuration. On average, {\scheme} can reduce the power usage by 16\% while sacrificing less than 0.01\% of system performance, thus leads to a 18.1\% of efficiency gain. This presents 80\% improvement achieved by the best dynamic scenario, in which we know ahead of time the optimal setting of each execution phase for each hardware component. Future research may focus on improving the efficacy of \scheme\ as well as extending the control of \scheme\ over more hardware resources to achieve more efficiency gain.

\bibliographystyle{plain}
\bibliography{main}

\begin{thebibliography}{10}

\bibitem{drg-cache}
A.~{Agarwal}, {Hai Li}, and K.~{Roy}.
\newblock Drg-cache: a data retention gated-ground cache for low power.
\newblock In {\em Proceedings 2002 Design Automation Conference}, 2002.

\bibitem{bala00}
Rajeev Balasubramonian, David Albonesi, Alper Buyuktosunoglu, and Sandhya
  Dwarkadas.
\newblock Memory hierarchy reconfiguration for energy and performance in
  general-purpose processor architectures.
\newblock In {\em Proceedings of the 33rd MICRO}, 2000.

\bibitem{bitirgen08}
Ramazan Bitirgen, Engin Ipek, and Jose~F. Martinez.
\newblock Coordinated management of multiple interacting resources in chip
  multiprocessors: A machine learning approach.
\newblock In {\em Proceedings of the 41st MICRO}, 2008.

\bibitem{eyeriss}
Y.~{Chen}, J.~{Emer}, and V.~{Sze}.
\newblock Eyeriss: A spatial architecture for energy-efficient dataflow for
  convolutional neural networks.
\newblock In {\em 2016 ACM/IEEE 43rd Annual ISCA}, pages 367--379, 2016.

\bibitem{choi06}
S.~{Choi} and D.~{Yeung}.
\newblock Learning-based smt processor resource distribution via hill-climbing.
\newblock In {\em 33rd ISCA}, page 239–251, Jun 2006.

\bibitem{imagenet}
J.~Deng, W.~Dong, R.~Socher, L.-J. Li, K.~Li, and L.~Fei-Fei.
\newblock {ImageNet: A Large-Scale Hierarchical Image Database}.
\newblock In {\em CVPR09}, 2009.

\bibitem{coscale}
Q.~{Deng}, D.~{Meisner}, A.~{Bhattacharjee}, T.~F. {Wenisch}, and
  R.~{Bianchini}.
\newblock Coscale: Coordinating cpu and memory system dvfs in server systems.
\newblock In {\em 2012 45th MICRO}, 2012.

\bibitem{dubach10}
C.~{Dubach}, T.~M. {Jones}, E.~V. {Bonilla}, and M.~F.~P. {O'Boyle}.
\newblock A predictive model for dynamic microarchitectural adaptivity control.
\newblock In {\em 2010 43rd MICRO}, Dec 2010.

\bibitem{hadi12}
Hadi Esmaeilzadeh, Adrian Sampson, Luis Ceze, and Doug Burger.
\newblock Neural acceleration for general-purpose approximate programs.
\newblock In {\em Proceedings of the 2012 45th MICRO}, 2012.

\bibitem{sysscale}
J.~{Haj-Yahya}, M.~{Alser}, J.~{Kim}, A.~G. {Yağlıkçı}, N.~{Vijaykumar},
  E.~{Rotem}, and O.~{Mutlu}.
\newblock Sysscale: Exploiting multi-domain dynamic voltage and frequency
  scaling for energy efficient mobile processors.
\newblock In {\em 2020 ACM/IEEE 47th Annual ISCA}, 2020.

\bibitem{lstm}
Sepp Hochreiter and Jurgen Schmidhuber.
\newblock Long short-term memory.
\newblock 9, 1997.

\bibitem{ipek08}
Engin Ipek, Onur Mutlu, Jos{\'e}~F. Mart\'{\i}nez, and Rich Caruana.
\newblock Self-optimizing memory controllers: A reinforcement learning
  approach.
\newblock In {\em Proceedings of the 35th Annual ISCA}, 2008.

\bibitem{cache-energy}
C.~{Isci}, A.~{Buyuktosunoglu}, C.~{Cher}, P.~{Bose}, and M.~{Martonosi}.
\newblock An analysis of efficient multi-core global power management policies:
  Maximizing performance for a given power budget.
\newblock In {\em 39th MICRO}, 2006.

\bibitem{mcpat}
S.~{Li}, H.~{Ann}, R.~D. {Strong}, J.~B. {Brockman}, D.~M. {Tullsen}, and N.~P.
  {Jouppi}.
\newblock Mcpat: An integrated power, area, and timing modeling framework for
  multicore and manycore architectures.
\newblock In {\em 2009 42nd MICRO}, Oct 2009.

\bibitem{caterpiller}
Y.~{Li} and A.~{Pedram}.
\newblock Caterpillar: Coarse grain reconfigurable architecture for
  accelerating the training of deep neural networks.
\newblock In {\em 2017 IEEE 28th International Conference on ASAP}, pages
  1--10, 2017.

\bibitem{vgg16}
S.~{Liu} and W.~{Deng}.
\newblock Very deep convolutional neural network based image classification
  using small training sample size.
\newblock In {\em 2015 3rd IAPR Asian Conference on Pattern Recognition
  (ACPR)}, 2015.

\bibitem{cache-vlsi}
A.~{Manan}.
\newblock Efficient 16 nm sram design for fpga’s.
\newblock In {\em 2018 5th International Conference on SPIN}, 2018.

\bibitem{samplepgo}
D.~{Novillo}.
\newblock Samplepgo - the power of profile guided optimizations without the
  usability burden.
\newblock In {\em 2014 LLVM Compiler Infrastructure in HPC}, page 22–28, Nov
  2014.

\bibitem{dvfs}
V.~Pallipadi and A.~Starikovskiy.
\newblock The ondemand governor.
\newblock 2006.

\bibitem{flicker}
Paula Petrica, Adam~M. Izraelevitz, David~H. Albonesi, and Christine~A.
  Shoemaker.
\newblock Flicker: A dynamically adaptive architecture for power limited
  multicore systems.
\newblock In {\em Proceedings of the 40th ISCA}, 2013.

\bibitem{dri-cache}
M.~{Powell}, {Se-Hyun Yang}, B.~{Falsafi}, K.~{Roy}, and N.~{Vijaykumar}.
\newblock Reducing leakage in a high-performance deep-submicron instruction
  cache.
\newblock {\em IEEE Transactions on VLSI Systems}, 2001.

\bibitem{beyondprofiling}
Chris Quackenbush and Mohamed Zahran.
\newblock Beyond profiling: Scaling profiling data usage to multiple
  applications.
\newblock {\em CoRR}, 2017.

\bibitem{charstar}
Gokul~Subramanian Ravi and Mikko~H. Lipasti.
\newblock Charstar: Clock hierarchy aware resource scaling in tiled
  architectures.
\newblock In {\em Proceedings of the 44th Annual ISCA}.

\bibitem{Reagen2016}
Brandon Reagen, Paul Whatmough, Robert Adolf, Saketh Rama, Hyunkwang Lee,
  Sae~Kyu Lee, José~Miguel Hernández-Lobato, Gu-Yeon Wei, and David Brooks.
\newblock Minerva: Enabling low-power, highly-accurate deep neural network
  accelerators.
\newblock In {\em ISCA}, 2016.

\bibitem{sherwood02}
Timothy Sherwood, Erez Perelman, Greg Hamerly, and Brad Calder.
\newblock Automatically characterizing large scale program behavior.
\newblock In {\em Proceedings of the 10th International Conference on
  ASPLOS-X}, 2002.

\bibitem{sherwood03}
Timothy Sherwood, Erez Perelman, Greg Hamerly, Suleyman Sair, and Brad Calder.
\newblock Discovering and exploiting program phases.
\newblock {\em IEEE Micro}, 2003.

\bibitem{tarsa19}
Stephen~J. Tarsa, Rangeen Basu~Roy Chowdhury, Julien Sebot, Gautham Chinya,
  Jayesh Gaur, Karthik Sankaranarayanan, Chit-Kwan Lin, Robert Chappell, Ronak
  Singhal, and Hong Wang.
\newblock Post-silicon cpu adaptation made practical using machine learning.
\newblock In {\em Proceedings of the 46th ISCA}, 2019.

\bibitem{multi2sim}
R.~{Ubal}, J.~{Sahuquilo}, S~{Petit}, and P.~{López}.
\newblock Multi2sim: A simulation framework to evaluate multicore-multithreaded
  processors.
\newblock In {\em 19th International Symposium on Computer Architecture and
  High Performance Computing}, pages 62--68, Oct 2007.

\bibitem{swhistory}
Wikipedia.
\newblock {History of Software}.
\newblock \url{https://en.wikipedia.org/wiki/History_of_software}.

\bibitem{wildstrom07}
Jonathan Wildstrom, Peter Stone, Emmett Witchel, and Michael Dahlin.
\newblock Machine learning for on-line hardware reconfiguration.
\newblock In {\em Proceedings of the 20th IJCAI, Hyderabad, India}, 2007.

\bibitem{cache-static}
A.~{Wiltgen}, K.~A. {Escobar}, A.~I. {Reis}, and R.~P. {Ribas}.
\newblock Power consumption analysis in static cmos gates.
\newblock In {\em 2013 26th Symposium on Integrated Circuits and Systems Design
  (SBCCI)}, pages 1--6, 2013.

\end{thebibliography}

\end{document}